\documentclass[11pt]{article}          % twocolumn
\usepackage{graphicx}
\usepackage{mathptmx} 

\usepackage[latin1]{inputenc}
\usepackage[english]{babel}

\usepackage{natbib}
\usepackage{amsfonts,amssymb,amsmath,amscd,latexsym}

\usepackage[a4paper,citecolor=blue,linkcolor=red,colorlinks=true]{hyperref}
\usepackage{appendix}

\usepackage{setspace} 
\usepackage[boxed]{algorithm}
\usepackage{algorithmic}
\usepackage{color}

\newtheorem{rem}{Remark}{\bf}{\rm}

\newcommand{\vs}{\vspace{0.5cm}}

\newcommand{\var}{\operatorname{Var}}

\newcommand{\eps}{\varepsilon}

\newcommand{\esp}{\mathbb{E}}

\newcommand{\btheta}{\boldsymbol{\theta}}
\newcommand{\bt}{\boldsymbol{\theta}}
\newcommand{\bz}{\mathbf{z}}
\newcommand{\bx}{\mathbf{x}}
\newcommand{\bxobs}{\mathbf{x}_\text{obs}}

\graphicspath{{images/}}
\usepackage{fullpage}

\begin{document}

\title{Efficient learning in ABC algorithms}

\author{Mohammed Sedki \\ 
Institut de Math\'ematiques et Mod\'elisation de Montpellier \\
Universit\'e Montpellier 2, France 
\and
Pierre Pudlo \\
Institut de Math\'ematiques et Mod\'elisation de Montpellier \\ 
Universit\'e Montpellier 2, France \& \\
Centre de Biologie et Gestion de Populations \\
INRA, Montpellier, France
\and
Jean-Michel Marin \\
Institut de Math\'ematiques et Mod\'elisation de Montpellier \\
Universit\'e Montpellier 2, France \\
\and
Christian P. Robert \\ 
CEREMADE, Universit\'e Paris Dauphine; France \& \\ 
CREST, INSEE, France
\and
Jean-Marie Cornuet \\
Centre de Biologie et Gestion de Populations \\  
INRA, Montpellier, France
}

\maketitle

\begin{abstract}
Approximate Bayesian Computation has been successfully used in population
genetics to bypass the calculation of the likelihood. These methods provide
accurate estimates of the posterior distribution by comparing the observed
dataset to a sample of datasets simulated from the model.  Although
parallelization is easily achieved, computation times for ensuring a
suitable approximation quality of the posterior distribution are still high.
To alleviate the computational burden, we propose an adaptive, sequential
algorithm that runs faster than other ABC algorithms but maintains
accuracy of the approximation. This proposal relies
on the sequential Monte Carlo sampler of \citet{DelMoral:Doucet:Jasra:2012}
but is calibrated to reduce the number of simulations from the model. The
paper concludes with numerical experiments on a toy example and on a
population genetic study of \textit{Apis mellifera}, where our algorithm was
shown to be faster than traditional ABC schemes.

\vspace{0.5cm}  \noindent Keywords: Likelihood-free sampler, Sequential Monte Carlo,
Approximate Bayesian computation, Population genetics
\end{abstract}

\newpage

\section{Introduction}
\label{sec:introduction}
In a parametric Bayesian setting, computation of the posterior
distribution of a parameter $\btheta\in\Theta\subset\mathbb{R}^p$
is a genuine issue when the likelihood $f\big(\bx|\btheta\big)$ is
not explicit. We denote by $\bx_{\text{obs}}\in\mathcal D$ the
observed data set, which not necessarily structured as an independent
and identically distributed (iid) sample. The probability density
function of the posterior distribution on $\btheta$ is 
\begin{equation}\label{eq:really.true.posterior}
  \pi\big(\btheta|\bx_{\text{obs}}\big)\propto
  \pi\big(\btheta\big)f\big(\bx_{\text{obs}}\big|\btheta\big).
\end{equation}
For complex models, including models with latent
processes, the computation of the likelihood function, namely $\btheta
\mapsto f\big(\bx_{\text{obs}} \big| \btheta \big)$, might be
expensive, or even impossible in a reasonable amount of time. Thus,
sampling from the posterior distribution is challenging. To bypass this
difficulty, the population genetics'community developped a technique based only on
the ability to simulate from the model, and it was popularised as
likelihood-free methods or approximate Bayesian computation (ABC).
We refer the reader to \cite{Marin:etal:2012} and \cite{Beaumont:2010} for surveys
on ABC methods and their uses in biology. For more
practical considerations, see e.g. \cite{Csillery:etal:2010}.

The ABC method is based on a sample, $(\bt_i, \bz_i)$, $i=1,2,\ldots$,
simulated from the joint distribution $\pi(\bt)f(\bz|\bt)$ on
$\Theta\times\mathcal D$. Then, the posterior density might be
estimated from the simulated sample, using a nonparametric density estimate
conditional on $\bz=\bxobs$. Since the data space often is
of high dimensionality, the estimation of the conditional
density is delicate (facing the curse of dimensionality). A solution
to this difficulty is that $\bxobs$ and the $\bz_i$'s are projected into a set of lower
dimension, say $d$, in a non-linear way, using some map $\eta:\mathcal
D\to\mathbb R^d$. The coordinates of $\eta(\bz)$ are often called the
summary statistics in the ABC literature. They correspond to the
following ersatz of the posterior:
\begin{equation}\label{eq:almost.true.posterior}
  \pi(\bt|\eta(\bxobs))\propto \pi(\bt)\int \mathbf 1\{\eta(\bz')=\eta(\bxobs)\} 
  f(\bz'|\bt)\, d\bz'. 
\end{equation}
As this paper is not about the selection of summary statistics, we will assume that
$\eta$ has been well chosen and that \eqref{eq:almost.true.posterior} is 
the quantity of interest in the absence of any more reliable approximation of
\eqref{eq:really.true.posterior}, see, \textit{e.g.},
\citet{Marin:etal:2012}. To simplify notations, we identify data sets
with their projections through $\eta$ onto $\mathbb R^d$. In other words,
$\bxobs$ and $\bz$ actually refer to $\eta(\bxobs)$ and $\eta(\bz)$
respectively, while the model distribution $f(\bz|\bt)$ and the posterior
distribution $\pi(\bt|\bxobs)$ refer to $\int \mathbf
1\{\eta(\bz')=\eta(\bz)\} f(\bz'|\bt)\, d\bz'$ and the distribution of
Eq.~\eqref{eq:almost.true.posterior} respectively.

To estimate the posterior, it is sufficient to produce a sample $(\bt_i,
\bz_i)$ localised around $\bxobs$, namely from
\begin{align}
  \pi_\epsilon(\bt, \bz) &:= \pi(\bt, \bz \,|\, d(\bz, \bxobs)\le\epsilon) \notag
  \\
  &\propto \label{eq:target-abc}
  \mathbf 1\{ d(\bz, \bxobs)\le\epsilon\} \pi(\bt)f(\bz|\bt),
\end{align}  
where $\epsilon$ is often called the tolerance level, see
\cite{Beaumont:2010}. It might be seen as a bandwidth of a
kernel density estimate with compact support kernel.  The
computational statistic literature has proposed some samplers from
this approximate distribution. The very first one is the likelihood-free
rejection sampler of \cite{Pritchard:etal:1999}.  As an alternative,
\cite{Marjoram:etal:2003} have introduced a Monte Carlo Markov chain
(MCMC) algorithm targeting \eqref{eq:target-abc} which does not
require any calculation of the likelihood either. Rejection sampling
and MCMC methods can be inefficient (in term of time complexity) when
the tolerance level $\varepsilon$ is small.  Thus various sequential
Monte Carlo algorithms \citep{Doucet:deFreitas:Gordon:2001,
  DelMoral:2004, Liu:2008} have been proposed, see
\cite{Sisson:Fan:Tanaka:2007}, \cite{Sisson:Fan:Tanaka:2009},
\cite{Beaumont:etal:2009}, \cite{Drovandi:Pettitt:2011} and
\cite{DelMoral:Doucet:Jasra:2012}. These algorithms start from a large
tolerance level $\epsilon_0$ and decrease the tuning parameter
over the iterations; thereby they gradually learn about the
target~\eqref{eq:target-abc}, \textit{i.e.}, in which part of the
parameter space $\btheta$'s should be simulated when given the
observation $\bx_\text{obs}$. All those schemes share the following
property: at the $t$-th iteration, they produce a Monte Carlo
sample distributed according to the ABC target with tolerance level
$\epsilon_t$, namely $\pi_{\epsilon_t}$, defined in
Eq.~\eqref{eq:target-abc}.

The algorithm of \cite{Beaumont:etal:2009} corrects the bias introduced
by \cite{Sisson:Fan:Tanaka:2007} (see also \cite{Sisson:Fan:Tanaka:2009})
and this is a particular Population Monte Carlo scheme
\citep{Cappe:etal:2004}. It requires fixing a sequence of decreasing
tolerance levels $\varepsilon_0 > \varepsilon_1 > \ldots >
\varepsilon_T$, which is not very realistic in practical problems.  In
contrast, the proposals of \cite{DelMoral:Doucet:Jasra:2012} and
\cite{Drovandi:Pettitt:2011} are adaptive likelihood-free versions of
the sequential Monte Car\-lo sampler \citep{DelMoral:Doucet:Jasra:2006}
and include a self-calibration mechanism for a sequence of
decreasing tolerance levels.

This paper proposes a different calibration mechanism of the
likelihood-free SMC sampler that
reduces the computation time when compared to the original
calibration scheme of \cite{DelMoral:Doucet:Jasra:2012}. Our proposal
assumes that the major proportion of the computational burden comes from
simulations from the model distribution. This typically holds for
complex models where ABC is one of the most used tools for Bayesian analysis, 
\textit{e.g.} for complex evolutionay scenarios in population
genetics. 

The plan of the paper is as follows. Start with presenting the
relevent earlier literature on ABC sampler in
Section~\ref{sec:background}: the accept-reject algorithm, the
MCMC-ABC sampler and a first sequential algorithm.  Throughout this
section, we detail properties of those algorithms that will prove useful
to understand our proposal, detailed in
Section~\ref{sec:efficient}. Following important remarks,
we expose the three stages of the proposed scheme: 
initialisation, the sequential part, and finally the stopping criterion on
iterations and post-processing. Section~\ref{sec:numerical}
is devoted to numerical experiments, and the paper concludes with a
discussion in Section~\ref{sec:discussion}.

\section{Background}
\label{sec:background}
We begin by connecting with the literature, and recall properties of
well-known methods that will prove useful to present our efficient
algorithm.

\subsection{The rejection sampler}

\begin{center}
\begin{minipage}{8cm}
\begin{algorithm}[H]
\caption{\label{algo:basic}\textbf{ABC rejection sampler}}
\begin{algorithmic}[1]
  \FOR {$i=1$ \textbf{to} $N_\text{prior}$}
	\STATE Generate $(\btheta_i, \bz_i)\sim \pi(\bt)f(\bz|\bt)$
        \STATE Compute $d_i=d(\bz_i,\, \bxobs)$
  \ENDFOR
  \RETURN the particles $(\bt_i,\bz_i)$ satisfying $d_i\le\varepsilon$
\end{algorithmic}
\end{algorithm}
\end{minipage}
\end{center}

\vs The sampler of \cite{Pritchard:etal:1999} targeting
\eqref{eq:target-abc} is detailed in Alg.~\ref{algo:basic}. The
algorithm simulates a number of $N_\text{prior}$ particles $(\bt_i,
\bz_i)$ from the joint distribution $\pi(\bt)f(\bz|\bt)$ and rejects
all particles that are too far from the observation. The size of the
output, say $N$ is of considerable importance: it impacts the
accuracy of the Monte Carlo estimate of any functional of the
target~\eqref{eq:target-abc}. Besides, we can see the
parameter $N_\text{prior}$ as representing the overall computational
effort, since the time complexity of the algorithm is $N_\text{prior}$,
when counted in the number of simulations from the model. 

Most often, a user has no clue about the tolerance level
$\varepsilon$ that should be used. One the other hand, one can readily set
the computational effort one is prepared to face, \textit{i.e.}, $N_\text{prior}$, and
set the size of the output, $N$, for the accuracy of the sample estimates. In
that perspective, $\epsilon$ appears as (a Monte Carlo
approximation of) the quantile of order $\alpha=N/N_\text{prior}$ of
the distance $d(\bz,\bxobs)$ when $\bz\sim \int \pi(\bt)f(\bz|\bt)\,
d\bt$.

\subsection{MCMC-ABC kernels}

\begin{center}
\begin{minipage}{10cm}
\begin{algorithm}[H]
\begin{algorithmic}[1]
  \STATE Generate some $(\btheta_0,\bz_0)$ 
  \FOR {$t=0$ \textbf{to} $T-1$}
    \STATE Generate $\btheta^\ast\sim \mathcal N(\btheta_t,\Sigma)$ \label{line:simule.btast}
    \STATE Generate $r\sim\mathcal U_{[0,1]}$ 
    \IF{ $r \leq \pi(\bt^\ast)/\pi(\bt_t)$ } 
      \STATE Generate $\bz^\ast \sim f\big(\bz|\btheta^\ast\big)$ \label{line:simule.bzast}
      \STATE \textbf{if} {$d(\bz^\ast, \bxobs)\le \epsilon$ } \textbf{then} $(\btheta_{t+1},\bz_{t+1})=(\btheta^\ast,\bz^\ast)$ 
      \textbf{else} $(\btheta_{t+1},\bz_{t+1})=(\btheta_t,\bz_t)$ \textbf{end if}
    \ELSE
      \STATE Set $(\btheta_{t+1},\bz_{t+1})=(\btheta_t,\bz_t)$
    \ENDIF
  \ENDFOR
  \RETURN the chain $(\btheta_t, \bz_t)_{t=0,\ldots,T}$
\end{algorithmic}
\caption{\label{algo:MCMC}\textbf{MCMC-ABC sampler}}
\end{algorithm}
\end{minipage}
\end{center}

\vs \citet{Marjoram:etal:2003} introduced a Markov chain Mon\-te
Carlo (MCMC) algorithm targeting \eqref{eq:target-abc} to avoid drawing
particles $(\bt, \bz)$ in low posterior probability areas. The main objective
was indeed to reduce the computational burden.  That chain representation is
a foundation stone of sequential ABC algorithms, and therefore of our proposals. Our
likelihood free MCMC algorithm is based on a Metropolis Hastings chain.
Given that the current state of the chain is $(\btheta_{t},\bz_t)$,
the proposal $\btheta^\ast$ is drawn from a Gaussian distribution centered at
$\btheta_t$ with covariance matrix $\Sigma$. The value of the new state
depends on a simulated data set ${\bz}^\ast$ from the likelihood
$f(\bz|{\btheta}^\ast)$. The new state $(\btheta_{t+1},\bz_{t+1})$ of the
chain is then equal to the proposal $(\btheta^\ast, \bz^\ast)$ with probability
\begin{equation}\label{eq:ratioMCMC}
\min\big\{1,\, {\pi(\btheta^\ast)}/{\pi(\btheta_t)} \big\}
\mathbf 1\big\{d({\bz}^\ast, \bxobs)\le \epsilon\big\}.
\end{equation}
Otherwise, the proposal is rejected and the state remains unchanged. The
complete MCMC scheme is detailed in Alg.~\ref{algo:MCMC}.

The invariant distribution of the above Markov chain is $\pi_{\epsilon}$ of
Eq.~\eqref{eq:target-abc}, see \citet{Marjoram:etal:2003}. However when the
tolerance level $\epsilon$ is small (and $\epsilon$ should be small), the Markov chain
has poor mixing properties. Indeed, in that case, the event $d(\bz^\ast, \bxobs)\le
\epsilon$ has small probability, and most of time the proposal is
rejected. Another caveat with this algorithm is the calibration of the proposal
covariance matrix $\Sigma$, which is a well known problem in MCMC. See, for instance,
\citet{robert:casella:2004}.

\begin{rem} \label{rem:ratio01}
If the prior distribution $\pi$ is uniform over
some compact set $\Theta_\text{prior}$ of the parameter space, the
ratio~\eqref{eq:ratioMCMC} is $1$ if the proposal $\btheta^\ast$ is in 
$\Theta_\text{prior}$ and $d({\bz}^\ast, \bxobs)\leq \epsilon$; otherwise, the ratio is $0$. 
\end{rem}
% \noindent \textbf{Moving according to $K_t$.}

% We propose to use  a Metropolis-Hastings kernel in step 4. For instance, we
% can compute the empirical variance $\tau_t^2$  of the
% $\big\{\btheta_i^t\big\}$'s in the set $\mathcal P_t$. For each
% particle $\big(\btheta^\ast_i,\bx^\ast_i\big)$ of $\mathcal P^\ast$,
% we generate some $\tilde\btheta$ according to the Gaussian distribution
% $\mathcal N\big(\btheta_i^\ast,2\tau_t^2\big)$ and some
% $\widetilde{\bx}$ according to the likelihood
% $f\big(\bx|\widetilde{\btheta}\big)$.  Since the prior is uniform, the acceptance rate for
% such a Gaussian proposal is binary, see
% \cite{DelMoral:Doucet:Jasra:2012} for more details.  If $\widetilde
% {\btheta}\in \Theta_\text{prior}$ and
% $d\Big(S\big(\widetilde{\bx}\big),S\big(\bx_{\text{obs}}\big)\Big) \le
% \varepsilon_{t+1}$, we set
% $\big(\btheta_i^{t+1},\bx_i^{t+1}\big)=\big(\widetilde{\btheta},
% \widetilde{\bx}\big)$. Otherwise, we do not move and set
% $\big(\btheta_i^{t+1},\bx_i^{t+1}\big)=\big(\btheta^\ast_i,\bx^\ast_i\big)$.

%\subsection{High performance computing}
\begin{rem} \label{rem:parallelMCMC} %
  The time complexity of the above MCMC scheme is at most $(T+1)$ because each
  iteration might require a new simulation from the model.  But, contrary
  to Alg.~\ref{algo:basic}, the method is no longer parallelisable because
  of the sequential nature of the Markov chain.
\end{rem}

\begin{rem} \label{rem:efficientMCMC} %
  In Alg.~\ref{algo:MCMC}, we have postponed the simulation of the proposed data
  set $\bz^\ast$ to line~\ref{line:simule.bzast}  instead of
  line~\ref{line:simule.btast} to save computational time. Indeed, in some cases,
  namely when $r \ge \pi(\bt^\ast)/\pi(\bt_t)$ the rejection of the proposal does
  not depend on the value of $\bz^\ast$.
\end{rem}

\begin{rem} \label{rem:normalMCMC} %
  In the above, generating the proposal $\bt^\ast$ from the Gaussian
  $\mathcal N(\bt_t,\Sigma)$ is an arbitrary choice. To generate
  $\bt^\ast$, any perturbation of $\bt_t$ of the form $\bt^\ast =
  \bt_t + \Delta$ is correct if distribution of $\Delta$, says
  $g(\cdot)$, is symmetric around $0$ (i.e., $g(\Delta)=g(-\Delta)$)
  and independent of $\bt_t$. If $g(\cdot)$ is not symmetric, the
  ratio~\eqref{eq:ratioMCMC} should be corrected by the factor
  $g(\bt^\ast-\bt_t)/g(\bt_t-\bt^\ast)$.
\end{rem}

\subsection{Sequential likelihood-free schemes}
\label{sec:state}

\begin{center}
\begin{minipage}{10cm}
\begin{algorithm}[H]
\caption{\label{algo:SMC}\textbf{Na\"{i}ve ABC-SMC sampler}}
\begin{algorithmic}[1]
  \FOR{$i=1$ \textbf{to} $N$}
     \STATE Generate $(\btheta_{0,i},\, \bz_{0,i}) \sim \pi(\btheta)f(\bz|\btheta)$
     \STATE Compute $d_{0,i}=d(\bz_{0,i},\bxobs)$
  \ENDFOR \label{line:endinit}
  \FOR {$t=0$ \textbf{to} $T-1$} \label{line:sequential}
     \STATE Sort the particles $(\btheta_{t,i}, \bz_{t,i})_{i=1,\ldots, N}$ according to their
           distances to the observation  $(d_{t,i})_{i=1,\ldots, N}$
     \STATE Set $\alpha_t$ such that $d_{t,i}\le \epsilon_{t+1}$ for all $i\le \alpha_t N$ \label{line:alpha}
%     \STATE Throw away each particle with index $i>\alpha_t N$ \label{line:reject}
%     \STATE Substitute this $t$-th array with  resampling of size $N$ from
%     the $\alpha_t N$ first particles  \label{line:resampling}
      \STATE Replace the particles $(\bt_{t,i},\bz_{t,i})_{i=1,\ldots, N}$
             by using residual resampling from $(\btheta_{t,i},\bz_{t,i})_{i=1,\ldots,\alpha_tN}$ \label{line:resampling}
      \FOR {$i=1$ \textbf{to} $N$}  \label{line:beginMarkov}
          \STATE Generate $(\btheta_{t+1,i},\bz_{t+1,i})$ applying one step of a
                 MCMC-ABC kernel to $(\btheta_{t,i},\bz_{t,i})$ with tolerance level
                 $\epsilon_{t+1}$
          \STATE Compute $d_{t+1,i}=d(\bz_{t+1,i},\bxobs)$
      \ENDFOR \label{line:endMarkov}
  \ENDFOR
  \RETURN the particles $(\bt_{T,i},\bz_{T,i})_{i=1,\ldots, N}$
\end{algorithmic}
\end{algorithm}
\end{minipage}
\end{center}

\vs \noindent The sequential Monte Carlo sampler of
\citet{DelMoral:Doucet:Jasra:2012} is the foundation stone of our efficient
algorithm. Assuming that \textit{(i)} the sequence of tolerance levels $\infty=\epsilon_0 > \epsilon_1 >
\epsilon_2 > \cdots > \epsilon_T$ is fixed, \textit{(ii)} resampling is
performed at each iteration and \textit{(iii)} the number of simulated data sets per particle,
denoted $M$ in the original paper, is $1$, we derive
\citet{DelMoral:Doucet:Jasra:2006,DelMoral:Doucet:Jasra:2012,delmoral:doucet:jasra:2012:resampling}
the validity of the algorithm. Each iteration reduces the tolerance level by accepting a proportion
$\alpha_t$ of the array, then fills the array by resampling among the
accepted particles and finally moves each particle by applying one step
of the Markov kernel described above.

This first sequential algorithm is given in Alg.~\ref{algo:SMC} and
can be explained as follow. At the end of
line~\ref{line:endinit}, the distribution of the particles provides an
approximation of the joint distribution denoted $\pi_{\epsilon_0}$ in
Eq.~\eqref{eq:target-abc} with $\epsilon_0=\infty$.
Line~\ref{line:sequential} sees the beginning of the sequential update
of the array. To understand this scheme, we assume that the
distribution of the particles in the array at the beginning is an
approximation of $\pi_{\epsilon_t}$ and justify that at the end of
line~\ref{line:endMarkov}, the particles in the new array provide an
approximation of $\pi_{\epsilon_{t+1}}$: 

\begin{itemize}
\item[1)] Because the conditioning events are nested, we have
  \[
  \pi_{\epsilon_{t+1}}(\btheta, \bz) = \pi_{\epsilon_{t}}\big(\btheta,\bz \,\big|\,
  d(\bz,\bx_\text{obs})\le \epsilon_{t+1}\big)
  \]
  and, at line \ref{line:resampling}, the $\alpha_t N$ first particles provide
  an approximation of $\pi_{\epsilon_{t+1}}$. Resampling from these
  $\alpha_t N$ particles does not modify this property. Hence,
  after the residual resampling step, the $N$ particles provide an approximation of $\pi_{\epsilon_{t+1}}$. 
\item[2)] At the end of line~\ref{line:endMarkov}, we have transformed the
  whole array through one step of a Markov kernel per particle,
  independently of the others. Since the invariant distribution of
  the chain is precisely $\pi_{\epsilon_{t+1}}$, the new array of
  particles still provides an approximation of $\pi_{\epsilon_{t+1}}$.
\end{itemize}

We recommand recoursing to residual resampling, which dominates the basic
multinomial resampling, see \citet{Douc:Cappe:Moulines:05}.
An important drawback of resampling is that it will introduce copies
of the particles. At transformation of the whole array through one step
of a Markov kernel aims at removing copies of the same particles
introduced by resampling. Indeed, we hope that different
copies will move in different directions. Sadly, however, particles modified
according to such a Metropolis-Hasting kernel have a non zero probability of
remaining at the same place. % This probability depends heavily on the
% value of the tolerance level $\epsilon_{t+1}$: if too small, almost any
% proposal will be refused.
%

The original sampler of \citet{DelMoral:Doucet:Jasra:2012} is much
more complex than the na\"{i}ve Alg.~\ref{algo:SMC}. It includes
the possibility of attaching several simulated data sets per particle, a
weighting of the particles, updated with the ratio in
Eq.~\eqref{eq:ratioMCMC}, a calibration scheme of the tolerance
level $\epsilon_t$ through time and a condition to decide wether
resampling should be applied at a given iteration.
However, their calibration scheme relies on a quantity named ``effective
sample size'' that does not take into account the repetitions
introduced by resamplings. 

Meanwhile, the calibration scheme of \cite{Drovandi:Pettitt:2011} involves a
na\"{i}ve strategy, calibrating the sequence of tolerance level $\epsilon_t$ such
that the quantile orders $\alpha_t$ at line~\ref{line:alpha} remain constant
over time, and might apply several steps of the MCMC kernel to remove the
copies, which is really time consuming for complex models. Moreover,
the number of steps might vary from one particle to the other to
ensure that each of the duplicates has really moved. 

\begin{rem} \label{rem:complexity.SMC} %
  Time complexity of this na\"{i}ve ABC-SMC sampler is at most
  $N\times(T+1)$. Indeed, running one step of the ABC-MCMC on a particle might
  require computation of a simulated data set.
\end{rem}

\begin{rem} \label{rem:efficiency.SMC} %
  To reduce the overall number of iterations $T$ and to reach the target
  with a low tolerance level as quickly as possible, the sequence
  ($\epsilon_t$) should decrease as fast as possible. But,
  resampling degrades the quality of the output (when compared with an iid
  sample of size $N$) by adding duplicates into the array. We still
  hope to obtain a final output without much duplicates because of the
  MCMC moves between lines~\ref{line:beginMarkov} and
  \ref{line:endMarkov}. This process actually removes some duplicates
  if some proposals $(\bt^\ast, \bz^\ast)$ are accepted. Hence, the
  $t$-th iteration does not add much duplicates into the array if a true move,
  \textit{i.e.}, acceptance, has non negligible
  probability. Obviously, this probability is large when
  $\epsilon_{t+1}$ is large. So to sum up, we have to balance between accuracy
  of the output (decreasing slowly the tolerance level) and reaching
  the target (decreasing quickly the tolerance level).
\end{rem}

\begin{rem} \label{rem:learn.SMC} %
  In order to be efficient, SMC schemes learn the posterior gradually. But, in
  the extrema case where the data set carries no information regarding the
  parameters, posterior and prior distribution are equal, and the effort to
  gradually learn the posterior is wasted. More
  generally, efficiency of SMC algorithms depends on the contrast between
  prior and posterior. See the toy example of Paragraph~\ref{sec:toy}.
\end{rem}

\begin{rem}
  At iteration $t$, the proposal variance $\Sigma$ within the ABC-MCMC kernel is typically
  set as twice the empirical variance of the particles at iteration $t-1$. Some numerical
  experimentations have supported the use of twice the empirical variances.  This scaling parameter
  is also used and justified within the importance sampling paradigm by \cite{Beaumont:etal:2009}.
\end{rem}

\section{An efficient self-calibrated algorithm}
\label{sec:efficient}
We propose here an improvement over the original ABC-SMC sampler. To
clarify the calibration issue, we step back to the na\"{i}ve sampler given in
Alg.~\ref{algo:SMC} that keeps the size of the current array constant over
time and that does not include any weighting of the particles. Moreover, we
take care of the repetitions inside the array to assess quality of the
current array.

Our goal is to decrease the tolerance level $\epsilon_{t+1}$ as much as
possible at each iteration, while keeping it sufficiently large so that the MCMC
moves between lines~\ref{line:beginMarkov} and \ref{line:endMarkov} of
Alg.~\ref{algo:SMC} remove a large part of the duplications introduced by
the resampling on line~\ref{line:resampling}.  We have also add an
initialisation stage that is able to decide whether or not it is possible to
learn from the target with a sequential scheme.

We assume that our final target is the ABC approximation
$\pi_{\epsilon}$ with a tolerance level $\epsilon$ that corresponds to a
pre-fixed quantile $\alpha$ of the distances between the observed
dataset $\bx_\text{obs}$ and a data set $\bz$ simulated from
$\int \pi(\btheta) f(\bz|\btheta)\, d\btheta$.

Some entries of the algorithm given below hinge upon a uniformly
distributed prior. Using variances to evaluate the difference between
the prior and a first rough estimate of the posterior makes no sense
if the prior is not flat. In addition, the ratio of acceptance of
the Metropolis Hastings Markov chain given in Eq.~\eqref{eq:ratioMCMC}
is either $0$ or $1$ in that case, see Remark~\ref{rem:ratio01}, which
simplifies the calibration scheme. With a uniform prior, no Metropolis
Hastings proposal is disadvantaged when $\bt^\ast$ falls into a
region of smallest prior probability. Of course, choosing a prior for
purely computational reasons is conceptually not very attractive. But
we can often reparameterize the model or find an explicit
transformation $\vec{\phi} = g(\bt)$ so that the desired prior on
$\bt$ leads to a uniform prior on $\vec{\phi}$. Then, any array of
particles in $(\vec{\phi}, \bz)$ returned by the algorithm might
easily transformed back into an array of particles in $(\bt, \bz)$.

\begin{rem}
  In rare situations, the reparametrization $g$ is not explicit. We
  refer the reader to Remarks~\ref{rem:init} and
  \ref{rem:nonuniform.MCMC} below to adapt the given methodology when
  the prior is not uniform.
\end{rem}

\subsection{The initialisation stage}\label{sub:init}

The initialisation stage provides us with a first rough estimate
of the target through an array of size $N$ with distribution
$\pi_{\epsilon_0}$.  The output of this stage will be the input of the
first iteration in the sequential stage. 
Our main goal is to detect whether or not there is more information contained
in the data set than in the prior. 
The algorithm detailed in Alg.~\ref{algo:init} is a modification of
the simple ABC rejection algorithm (Alg.~\ref{algo:basic}) gradually
increasing the number of particles to decrease the tolerance
level. More precisely, when compared with Alg.~\ref{algo:basic},
we fix the size of the output, say $N$, and increase the total
number of simulations (denoted $N_{prior}$ in Alg.\ref{algo:basic}) by
steps of size $N$. Adding $N$ new particles to the array is performed
between lines~\ref{line:begin.add} and \ref{line:end.add}. If the
initialisation stage is stopped with the current value of $K$, then the
best $N$ particles are returned. They correspond to a tolerance level
$\epsilon_0$ which is computed at line~\ref{line:eps0}.  

To check whether or not the current approximation of the target has learned
anything when compared with the prior, we propose to compare the determinant
of the variance of the current array with the one of the prior. If the true
posterior is unimodal, then the determinant of its variance is lower
than the one of the flat prior.  At step $K$, the gap between the
prior and the current approximation of the posterior (given by the $N$
first particles of the array) is then measured with the
difference between the determinants of their variances. In other
words, the concentration of the output is measured through the determinant of
its covariance matrix. Thus, at line~\ref{line:delta} of
Alg.~\ref{algo:init}, we compute this determinant $v_K$ as if the
output was made of the first $N$ particles among the array of $KN$ particles.
We stop the loop at the first tolerance level
$\epsilon_0$ for which the determinant of the variance matrix is twice
smaller than the determinant of the variance matrix of the prior. If,
we reach the tolerance level of the final target, we stop
the initialisation stage, as well as the whole procedure at that
level, and return the $N$ first particles of the array as final output
of the whole scheme.

\begin{center}
\begin{minipage}{15.5cm}
\begin{algorithm}[H]
\caption{\label{algo:init}\textbf{Initialisation of the efficient
    self-calibrated algorithm}}
\begin{algorithmic}[1]
  \FOR{$i=1$ \textbf{to} $N$}
    \STATE Generate $(\btheta_i,\bz_i) \sim  \pi(\btheta)f(\bz|\btheta)$
    \STATE Compute $d_i = d\big(\bz_i,\bx_{\text{obs}}\big)$
  \ENDFOR   
    \STATE Sort the particles $(\bt_i,\bz_i)_{i=1,\ldots,N}$ according to their distances
           to the observation $(d_i)_{i=1,\ldots,N}$
    \STATE Set $v_1=\det\big(\var(\btheta_{1},\ldots,\bt_{N})\big)$ \label{line:delta.prior}
    \STATE Set $K=1$ and $\epsilon_0=\infty$
  \WHILE {$\epsilon_0\ge \epsilon$ \textbf{and} $v_K\ge v_1/2$} \label{line:stop}
    \STATE Set $K = K + 1$ 
    \STATE {$\#$ \tt Generate $N$ new particles:}
    \FOR{ $i=(K-1)N +1$ \textbf{to} $KN$} \label{line:begin.add}
      \STATE Generate  $(\btheta_i,\bz_i)\sim \pi(\btheta)f(\bz|\btheta)$ 
      \STATE Compute $d_i=d\big(\bz_i,\bx_{\text{obs}}\big)$
    \ENDFOR \label{line:end.add}
   \STATE Sort the particles $(\bt_i,\bz_i)_{i=1,\ldots,KN}$ according to their distances
          to the observation $(d_i)_{i=1,\ldots,KN}$
   \STATE Set $v_K=\det\big(\var(\btheta_{1},\ldots,\bt_{N})\big)$ \label{line:delta}
   \STATE Update $\epsilon_0=d_N$ \label{line:eps0}
  \ENDWHILE
  \RETURN the particles $(\bt_i,\bz_i)_{i=1,\ldots,N}$
\end{algorithmic}
\end{algorithm}
\end{minipage}
\end{center}

The aim of this first stage is to seek a good starting point for the
sequential stage. It might be useless at first sight, but it indeed offers
the two following advantages.  First, it initializes the sequential stage with
a first approximation of the target. Second, it prevents the user
from running the sequential algorithm if not efficient, \textit{i.e.}, if we
are in cases comparable with the ones of Remark~\ref{rem:learn.SMC} above.

\begin{rem}\label{rem:init}
  Comparing the prior and a rough estimate of the posterior through the
  determinant of the variance matrices might be inappropriate in some
  situations. It is efficient only when the prior is flat and
  the posterior is unimodal. The one half factor in the criterion was
  set quite arbitrarily and may be changed.
  \\
  If the user has no clue as to which criterion to use to stop the loop in
  this initialisation stage, he or she can graphically represent the prior
  and a kernel density estimate of the current $N$ first particles to
  detect if they have sufficiently diverged, or build an criterion
  adapted to the context.
\end{rem}

\subsection{The sequential stage}

To sum up Remark~\ref{rem:efficiency.SMC} above,
we have to balance accuracy of the output (decreasing slowly the
tolerance level) and quickly reaching the target (decreasing quickly
the tolerance level) in the calibration scheme. To fathom the
efficiency of the Markov steps in removing duplicates, let us denote
by $n_t$ the number of distinct particles in the array at the beginning of
the $t$-th iteration of Alg.~\ref{algo:SMC}.  We assume that
\textit{(i)} the proportion $\alpha_t$ computed at line~\ref{line:alpha} is
small, \textit{i.e.}, $\alpha_t\ll 1$, \textit{(ii)} $\alpha_tN$ is an
integer and $N$ is a multiple of $\alpha_tN$ and \textit{(iii)}
resampling in line~\ref{line:resampling} is residual resampling,
\textit{i.e.}, the array is made of $1/\alpha_t$ identical copies of
the $(\alpha_t N)$ surviving particles. Then,
\begin{equation}\label{eq:n_t}
\esp(n_{t+1})\ge \big(\alpha_t+\rho_t+o(\alpha_t)\big)\esp(n_t),
\end{equation}
where $\rho_t$ is the probability that, within one step of
the MCMC-ABC kernel at tolerance level $\epsilon_{t+1}$, the particle
has moved somewhere else. The proof of the first order approximation in
Eq.~\eqref{eq:n_t} is given in the Appendix.

\begin{algorithm}
\caption{\textbf{Iteration $t$ of the self-calibrated sampler}\label{algo:gas}}
%\algsetup{indent=2em}
\begin{algorithmic}[1]
  \STATE Sort the particles $(\btheta_{t,i},\bz_{t,i})_{i=1,\ldots,N}$ according to their
  distances to the observation $(d_{t,i})_{i=1,\ldots,N}$\\[.5em]
    {$\sharp$ \texttt{Calibration of $\alpha$} }
    \STATE Set $\alpha = 0.0$
    \REPEAT \label{line:begin.cal}
      \STATE Set $\alpha' = \alpha + 0.01$ \label{line:increase.alpha}
      \STATE Set $\varepsilon' = d_{t,\alpha' N}$ \label{line:compute.epsilon}
      \FOR {$i=\alpha N+1$ \textbf{to} $\alpha' N$} \label{line:begin.proposals}
          \STATE Generate $\bt_i^\ast\sim \mathcal N(\btheta_{t,i},
          \Sigma)$
          and $\bz_i^\ast\sim
          f(\bz|\bt_i^\ast)$ \label{line:move0}
     \ENDFOR \label{line:end.proposals}
     \STATE Set $N_\text{move}$ equal to the number of
     $(\btheta_i^\ast, \bz_i^\ast)_{i=1,\ldots,\alpha'N}$ satisfying \label{line:Nmove}
     $\btheta_i^\ast \in \Theta$ and $d(\bz_i^\ast,\bxobs)\le \varepsilon'$
     \STATE Update $\rho = N_\text{move}/(\alpha' N)$ and $\alpha =
     \alpha'$ \label{line:estimate.rho}
   \UNTIL {$\alpha + \rho \ge 1.0$} 
   \STATE Set $\alpha_t = \alpha$ and $\varepsilon_t = \varepsilon'$ \label{line:end.cal}
   \\[.5em]
   { $\sharp$ \texttt{Computation of the new array} }
   \FOR {$i=1$ \textbf{to} $\alpha_t N$} \label{line:begin.part1}
   \STATE \textbf{if} {$\bt_i^\ast \in \Theta$ \textbf{and}
     $d(\bz_i,\bxobs)\le \varepsilon_{t}$} 
   \textbf{then}
     $(\btheta_{t+1, i},\, \bz_{t+1,\, j})=(\btheta_j^\ast, \bz_j^\ast)$
   \textbf{else}
     $(\btheta_{t+1, i},\, \bz_{t+1,\, j})=(\btheta_{t, i},\, \bz_{t,\, j})$
   \textbf{end if}
   \ENDFOR \label{line:end.part1}
   \STATE  \label{line:begin.part2} Replace the particles $(\bt_{t,i},\bz_{t,i})_{i=1,\ldots, N}$
             by using residual resampling from $(\btheta_{t,i},\bz_{t,i})_{i=1,\ldots,\alpha_tN}$ \label{line:pickup}
   \FOR {$i=\alpha_{t}N + 1$ \textbf{to} $N$} 
   \STATE Generate $\btheta^\ast \sim \mathcal N(\btheta_{t, i}, \Sigma)$ 
   and $\bz^\ast \sim f(\bz|\btheta^\ast)$ \label{line:move1}
   \STATE \textbf{if} {$\bt^\ast \in \Theta$ \textbf{and}
     $d(\bz^\ast,\bxobs)\le \varepsilon_{t}$} 
   \textbf{then}
     $(\btheta_{t+1,\, i}, \bz_{t+1,\, i})=(\bt^\ast, \bz^\ast)$
   \textbf{else}
     $(\btheta_{t+1,\, i}, \bz_{t+1,\, i})=(\bt_{t,\, i}, \bz_{t,\, i})$
   \textbf{end if} \label{line:move2}
   \ENDFOR \label{line:end.part2}
\end{algorithmic}
\end{algorithm}

Once a value of $\epsilon_{t+1}$ is chosen, $\alpha_t$ and $\rho_t$ follow.
The calibration of $\epsilon_{t+1}$ that we present
here actually rely on calibrating $\alpha_t$. We seek $\alpha_t$
to be as small as possible (for speed reasons), but such that $\alpha_t+\rho_t$
remains large for accuracy because of Eq.~\eqref{eq:n_t} (even when this
assumption $\alpha_t \ll 1$ does not hold). The largest value of
$\alpha_t+\rho_t$ we can reach for sure is $1$ (by setting
$\alpha_t=1$). Hence we adopt the rule to search for the
smallest $\alpha_t$ such that $\alpha_t+\rho_t=1$.
Another perspective on this calibration rule is one of a simple
trade-off between speed (low values of $\alpha_t$) and accuracy
(large values of $\rho_t$).

The main difficulty with this calibration rule is that we have no
explicit formula for $\rho_t$, which should thus be estimated while
performing the calibration of $\alpha_t$. But we want to abstain from
generating more simulations from the model than necessary (because of their
computational cost). This certainly is the reason why the
iterations of the self-calibrated sampler we propose in
Alg.~\ref{algo:gas} is so complex. It however relies on a simple fact: if we
start from a given particle $(\bt, \bz)$ and want to run one step of
the ABC-MCMC kernel, we first have to generate a proposal $(\bt^\ast,
\bz^\ast)$ and afterward decide whether or not it is accepted. Computing the
proposals is time expansive (because of the simulated $\bz^\ast$),
but it does not depend on the current tolerance level. Hence if we want
to change the tolerance level of the chain, we do not need to generate
another proposal but we only have to correct the decision whether or not to
accept this proposal.

The other fact on which the calibration scheme relies is that, when
performing residual resampling to obtain an array of size $N$ from
a smaller array of size $\alpha N$, the result begins with one
identical copy from the small array. When $\alpha$ increases, this first
copy of the small array is increased.

We are now ready to explain Alg.~\ref{algo:gas} in full detail. We
calibrate $\alpha_t$ between lines~\ref{line:begin.cal} and
\ref{line:end.cal}, fill the first part of the new array
(corresponding to the first identical copy of the remaining particles)
between lines~\ref{line:begin.part1} and \ref{line:end.part1} and
at last compute the rest of the new array between
lines~\ref{line:begin.part2} and \ref{line:end.part2}.

The detailed description of the calibration stage is as follows.
At each passage through the repeat-loop, $\alpha'$ is increased
(line~\ref{line:increase.alpha}) when compared with the old value
$\alpha$, the current tolerance level $\epsilon'$ is updated
(line~\ref{line:compute.epsilon}), and one should think of the
$\alpha' N$ first particles as the remaining particles at level
$\epsilon'$. Proposals towards the step of the ABC-MCMC
kernel are computed between lines~\ref{line:begin.proposals} and
\ref{line:end.proposals}. Note that a large part of the proposals we
store for future use are already computed: the newly made particles
have indices varying from $\alpha N +1$ to $\alpha'N$. We then calculate
$N_\text{move}$ the number of proposals among the first $\alpha' N$ ones that would be accepted if
we were performing one iteration of ABC-MCMC scheme with tolerance
level $\epsilon'$ (line~\ref{line:Nmove}). And the
current value of $\rho_t$, the probability of moving somewhere else
in one iteration of the ABC-MCMC scheme with toleranace level $\epsilon'$, is estimated
with the proportion of accepted proposals, namely $N_\text{move}/(\alpha' N)$, at
line~\ref{line:estimate.rho}. 

Once $\alpha_t$ is fixed, the first part of the new array (from index
$1$ to $\alpha_tN$) is almost computed. Indeed, we only have to
store the decisions whether or not the proposals of the ABC-MCMC kernel are
accepted (lines~\ref{line:begin.part1}--\ref{line:end.part1}). We end
up performing resampling from the remaining particles
(line~\ref{line:pickup}) and apply one step of the chain with level
$\epsilon'$ (line~\ref{line:move1}) until the new array is completely
filled.

\begin{rem} %
  To alleviate the notations in Alg.~\ref{algo:gas}, we have
  considered that $\alpha N$ and $\alpha' N$ are integer numbers. If
  not, $\alpha N$ and $\alpha' N$ should be replaced with their
  integer parts.
\end{rem}

\begin{rem} \label{rem:nonuniform.MCMC} %
  If the prior is not uniform, the acceptance stage of the Metropolis
  Hastings proposal $\bt_i^\ast$ at lines~\ref{line:Nmove} and
  \ref{line:move2} should be corrected to fit with the acceptance
  probability of Eq.~\eqref{eq:ratioMCMC}. Then, as in
  Alg.~\ref{algo:MCMC}, the decision is taken with the help of a random
  $r_i\mathcal U_{[0,1]}$ that should be computed
  together with the proposal, i.e. at lines~\ref{line:move0} and
  \ref{line:move1}.
\end{rem}

% \begin{rem} \label{rem:bias}%
%   \textbf{XXX a completer serieusement! XXX}\\
%   Bias introduced by the calibration scheme?  Negligible because
%   $\alpha_t$, $\rho_t$ and $\epsilon_{t+1}$ are estimates computed on
%   large samples. Would require a more formal proof, but it is another
%   job than explaining the rationale behind the algorithm (proving that
%   self-calibrated algorithms, including resampling stages, if often a
%   very complex problem).
% \end{rem}

\subsection{Stop criterion and post-processing}
\label{sub:stop}

We include a criterion towards stopping the sequential stage. Most often,
at each iteration $t$, $\alpha_{t+1}\ge \alpha_{t}$ and $\rho_{t+1}\le
\rho_{t}$. This implies that it becomes more and more difficult to
remove the duplicates introduced by the resampling stage. In that case the
Markovian moves are needlessly time expensive. Hence, after some iterations, it would be better to stop
the SMC stage and perform (if needed) a simple rejection step to reduce the
tolerance level and reach the value of the target.  Pratically, we
recommand to stop when $\rho_t\le 0.1$.
% If by chance,
% we reach the tolerance level $\epsilon$ of the final target
% before, the whole algorithm is stopped and returns the current array
% of particles.  Otherwise (and it has been always the case in our
% simulation studies), the last tolerance level $\epsilon_T$ of the
% sequential stage is larger than our target. We then have to perform a
% last rejection step that filter particles with data sets too far from
% the observation $\bxobs$.

% \begin{rem} %
%   We shall also note here that we can still apply Beaumont REF's local
%   regression at the end to improve the accurary of the output. XXX A
%   completer XXX
% \end{rem}

\begin{rem} \label{rem:hpc2} %
  One of the major appeals of the classical ABC algorithm is its
  ability to be parallelized on multicore architectures and computing
  clusters. In our proposal, the many simulations required by the algorithm 
  can be computed in parallel. However, we need a
  memory shared architecture to run the sequential algorithm, since each
  round through the loop depends on the previous simulated sample. If
  the steps of simulating a dataset from the model and computing the summary
  statistics are not fast, we observe good parallel program performance
  on a multicore computer using the OpenMP API (see
  \url{http://openmp.org}). 
  
  If one wants to parallelize the program on several computers of a
  cluster, the situation is quite different. We refer the reader to
  \citet{marin:pudlo:sedki:2013}.
\end{rem}

\newpage

\section{Numerical experiments}%\label{sec:num}
\label{sec:numerical}
We have implemented our proposal in two numerical experiments. The first
is a toy example, presented in Sec.~\ref{sec:toy}, which is simple
enough to compare the quality of the output with the true posterior
and to study its average behaviour through independent replicates of the whole
scheme.  The population genetics experiment developped in
Sec.~\ref{sec:popgen} is a phylogeographic analysis of the European
honeybee (\textit{Apis Mellifera}), which serves as an example of
complex models evaluated with ABC. In that last experiment, the dimension
of the parameter is eight, and the dimension of the data set (ie the number of
summary statistics) is thirty. 

\subsection{Evaluating accuracy and speed}\label{sec:measurements}

To study the numerical performances our proposal, we need to quantify the accuracy of
the final sample and its efficiency.

\noindent{\bf $\triangleright{}$ Accuracy of the final sample.}
The accuracy of the output, seen as an array of weighted
particles is measured with the effective sample size (ESS). We calculate this ESS by
considering only distinct particles with their aggregated weight,
as in \cite{schafer:2013}. If one particle $(\bt,\bz)$ is repeated $k$
times with weights $\omega_{i_1}, \ldots, \omega_{i_k}$, its weight
after agregation is $\omega_{i_1}+ \cdots + \omega_{i_k}$. Assuming
the aggregation has been performed, this quality indicator is
\begin{equation} \label{eq:ESS}
  ESS = \left(\sum_i \omega_i\right)^2 \bigg/ 
  \left(\sum_i \omega_i^2 \right),
\end{equation}
where $\omega_i$ is the weight of the $i$-th particle in the
aggregated array. In many cases, the array is
obtained by a resampling step followed by a systematic assignment of
the weight $1/N$ to each particle. Such an array might be composed of
a small number of distinct particles repeated many times. If
we apply directly the formula \eqref{eq:ESS} to such arrays, the ESS would
be equal to $N$, which is not a fair assessment of the quality. 

The ESS gives a pessimistic estimate of the quality of a weighted array
in terms of variance, see \citet[Chapter~2]{Liu:2008}. More precisely,
if the original weighted array is of size $N$, and if the particles
are independent, $ESS/N$ is a lower bound of the ratio comparing
the variance of empirical averages over the weighted array and the variance
of empirical averages of an array of size $N$ whose entries are iid from
the target distribution. 

Relying on ESS to assess quality of the output can be criticized
for two reasons. First, our proposals are seeking to maintain a
fixed ESS, potentially without maintaing a fixed accuracy with respect to the
target. Second, our particles are dependent due to the
resampling. However a numerical analysis was performed on the toy
example provided below to study the relevance of the ESS, and the number of
distinct particles, as measures of accuracy. It appears that this
criterion is quite fair, see below.

\noindent{\bf $\triangleright{}$ Efficiency of the proposal.} To
quantify the efficiency of our scheme against the rejection
algorithm, we introduce an index number $r$, called the
gain factor, that is defined as follows. If when
targeting a given level $\epsilon$, the whole algorithm has required
$n_\text{model}$ simulations from the model, denote by
$n_\text{final}$ the ESS of the output. We then count
$n^0_\text{model}$, the number of simulations required by the rejection
algorithm to reach the same level $\epsilon$ and to produce an output of
$n_\text{final}$ particles. The gain factor is then defined as
\begin{equation} \label{eq:gain_factor}
r = n^0_\text{model} / n_\text{model}.
\end{equation}
If the model is complex enough, ABC samplers spend the major part of the
computing time simulating from the model. Associated times requires to sort arrays,
to resample or to generate the $\btheta$'s are negligible. Hence $n^0_\text{model}$ and
$n_\text{model}$ are approximations of the computation times of the
rejection algorithm and of our scheme, respectively. Both algorithms are
tuned to produce an output with the same accuracy (measured though the ESS
and the final tolerance level). Thus, the gain factor may be seen as
an approximation of the time factor we save using our scheme
for a complex model whose simulations require significant time.

\subsection{A toy example}\label{sec:toy}

\begin{figure}[n]
  \begin{tabular}{cc} \small
    Mean of the target: & \small Median of the target:\\[-.8cm]
    \includegraphics[width=.45\linewidth]{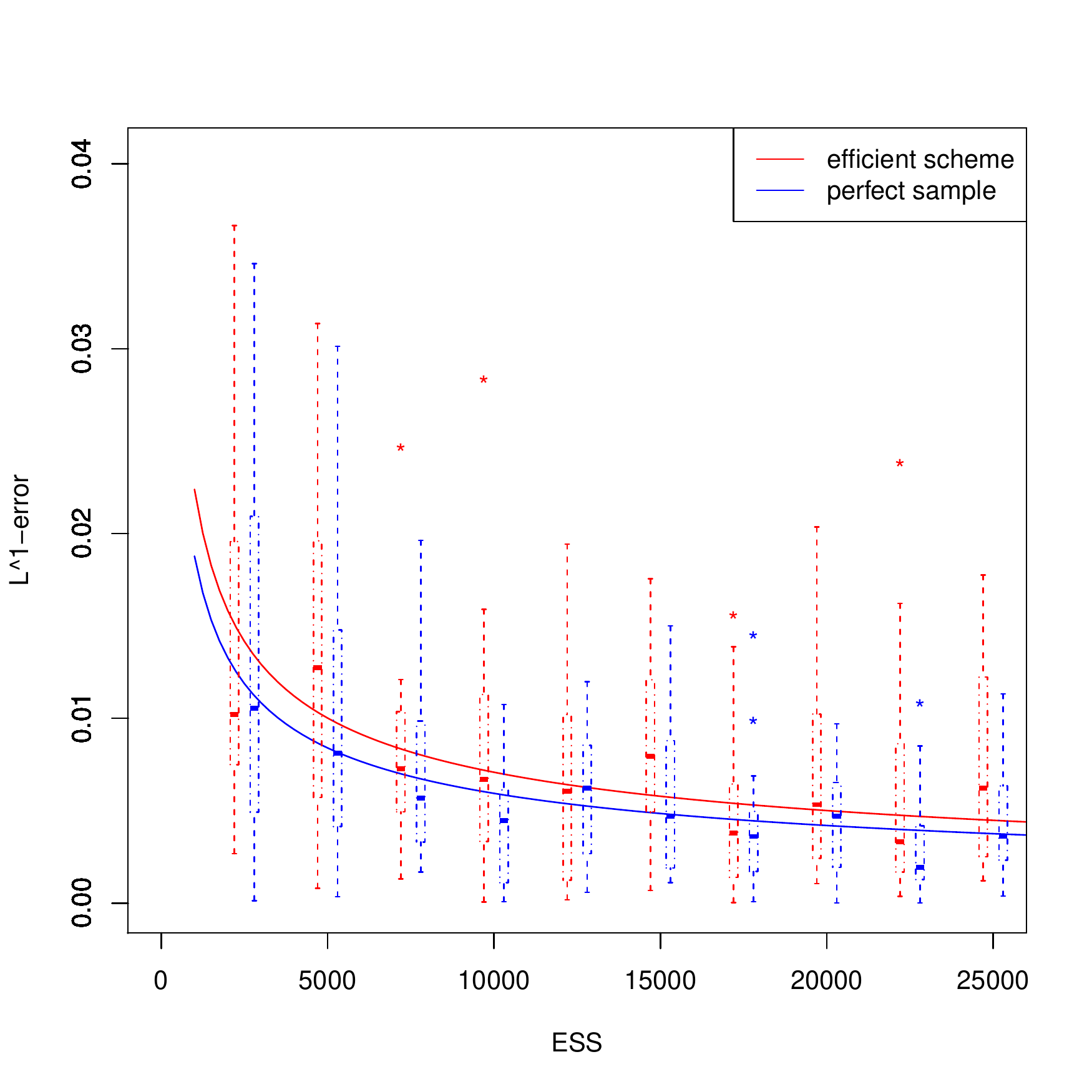}&
    \includegraphics[width=.45\linewidth]{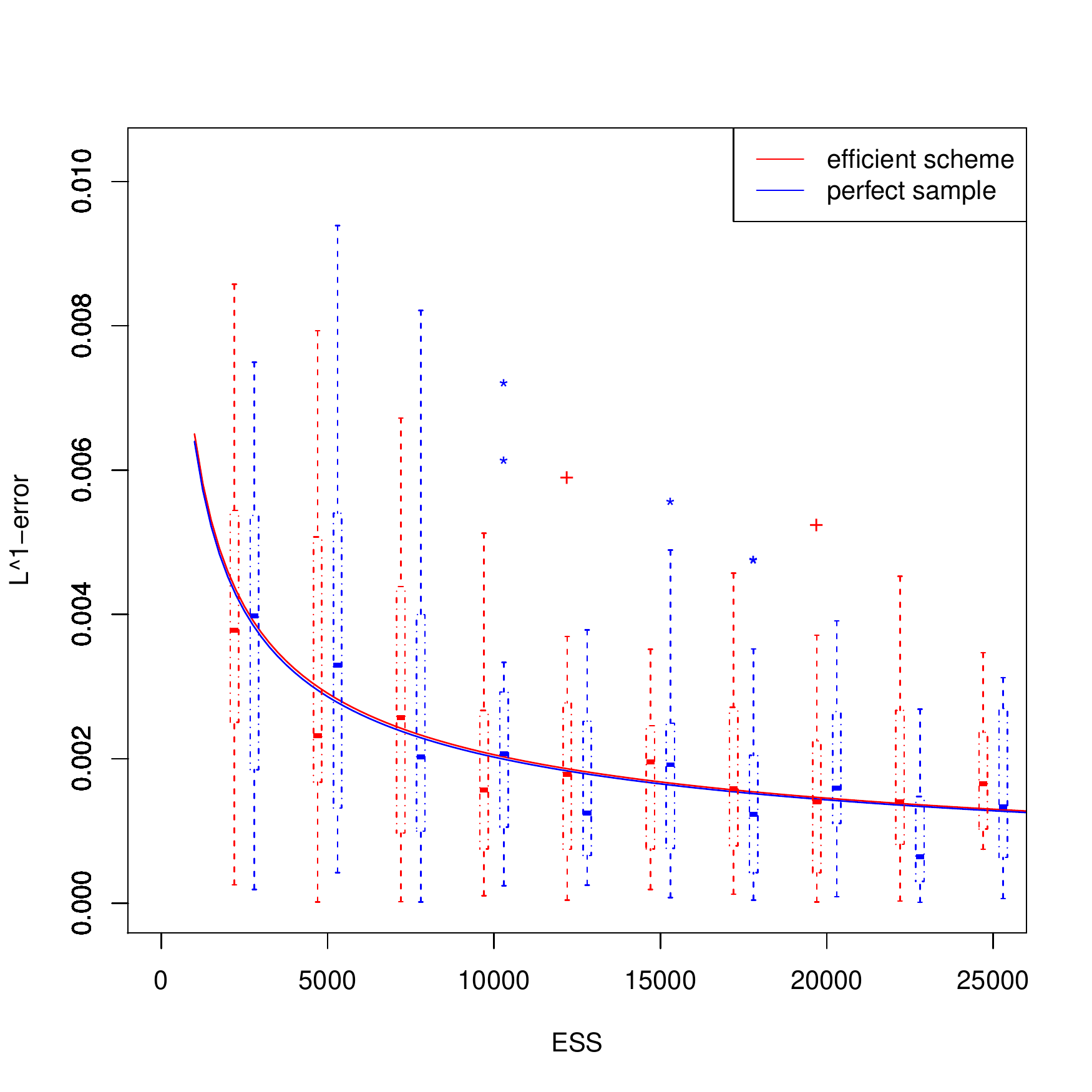}\\
    \small $1$st quartile of the target: & 
    \small $3$rd quartile of the target:\\[-.8cm]
    \includegraphics[width=.45\linewidth]{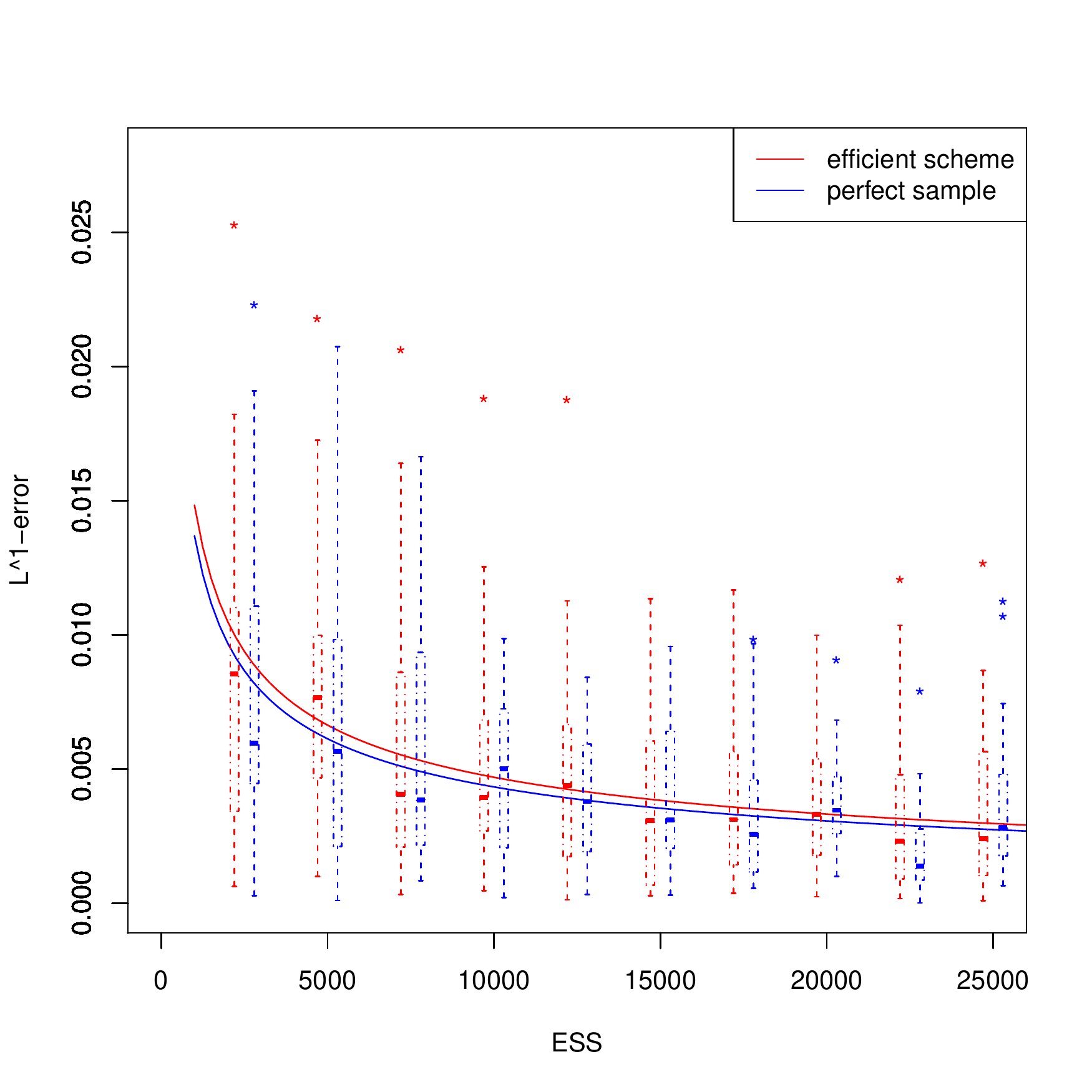}&
    \includegraphics[width=.45\linewidth]{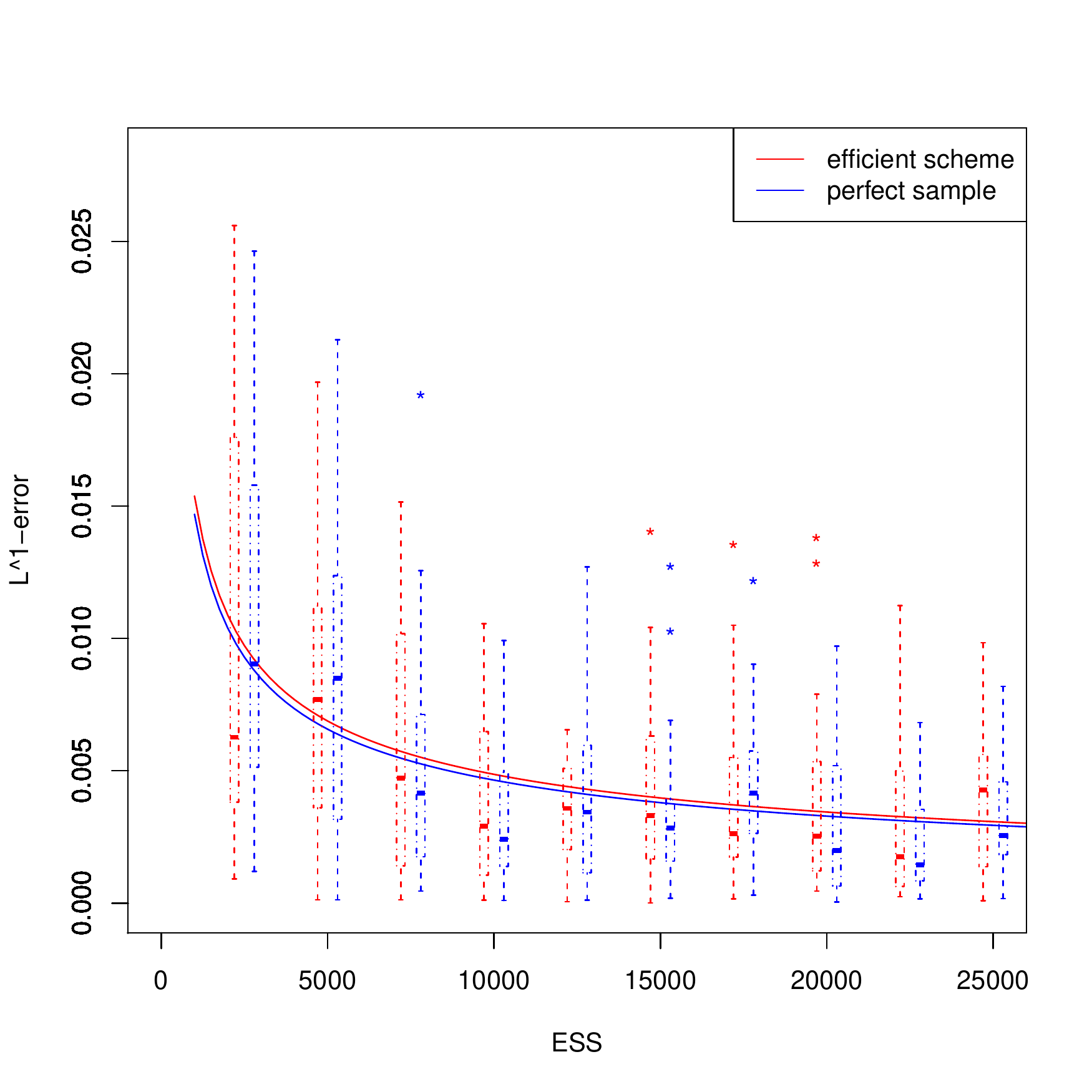}\\
  \end{tabular}
  \caption{\label{fig:good.ESS} {\bf Comparison between errors} associated to our self-calibrated and to the rejection algorithms.
    We estimate four functionals of the target: the mean (\textit{top left}), the median (\textit{top
    right)}, the $1$st and the $3$rd quartile (\textit{bottom left
    and right}). Each boxplot represents the variation of the $L^1$-error over $20$
    independent replicates for values of ESS$\approx 2\,500,\ 5\,000,\ldots,\ 25\,000$. 
    The continuous curves result from a linear fit of the $L^1$-error against $1/\sqrt{ESS}$ over
    the whole 200 independent runs.
  }  
\end{figure}

\begin{figure}[n]
    \includegraphics[width=0.45\linewidth]{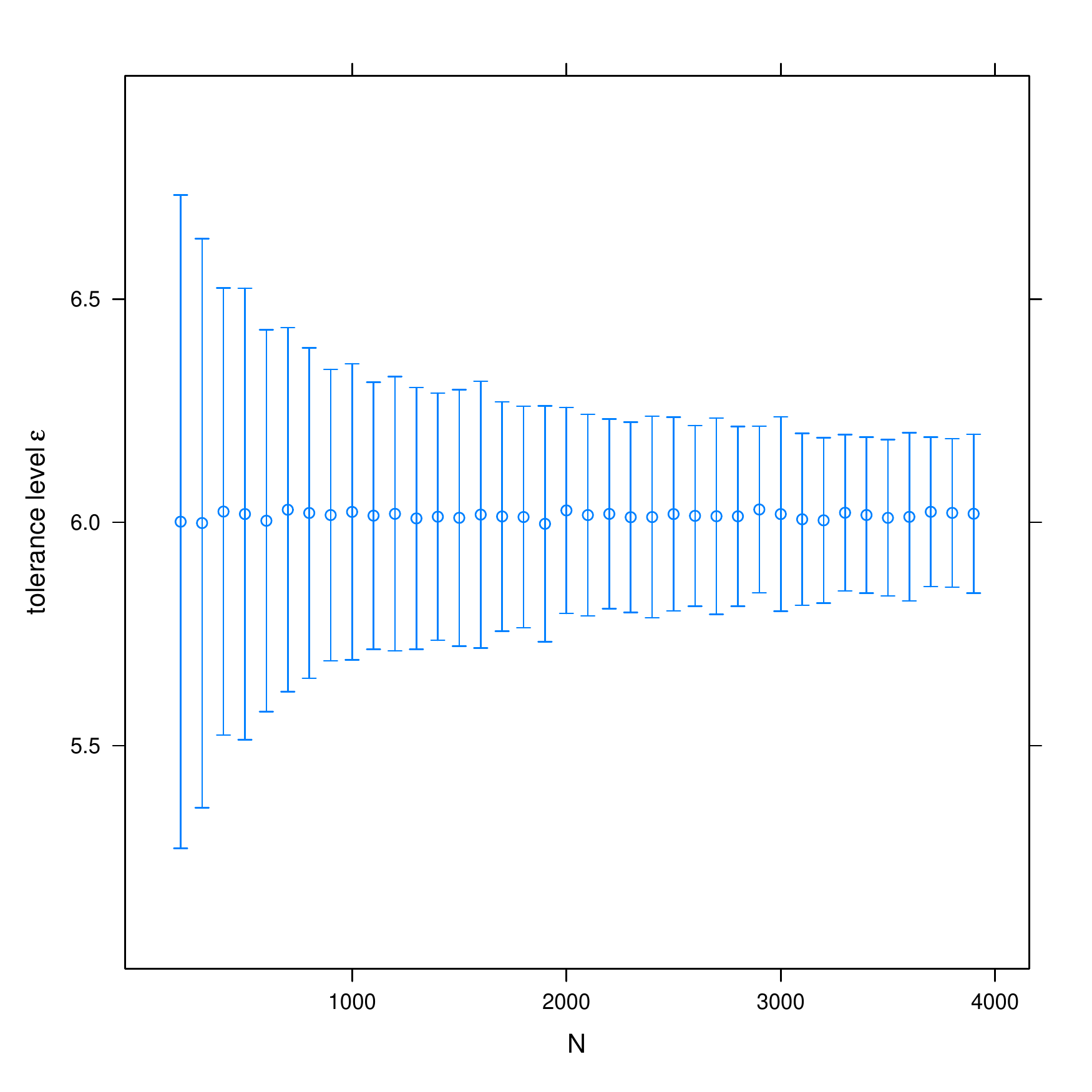}
    \includegraphics[width=0.45\linewidth]{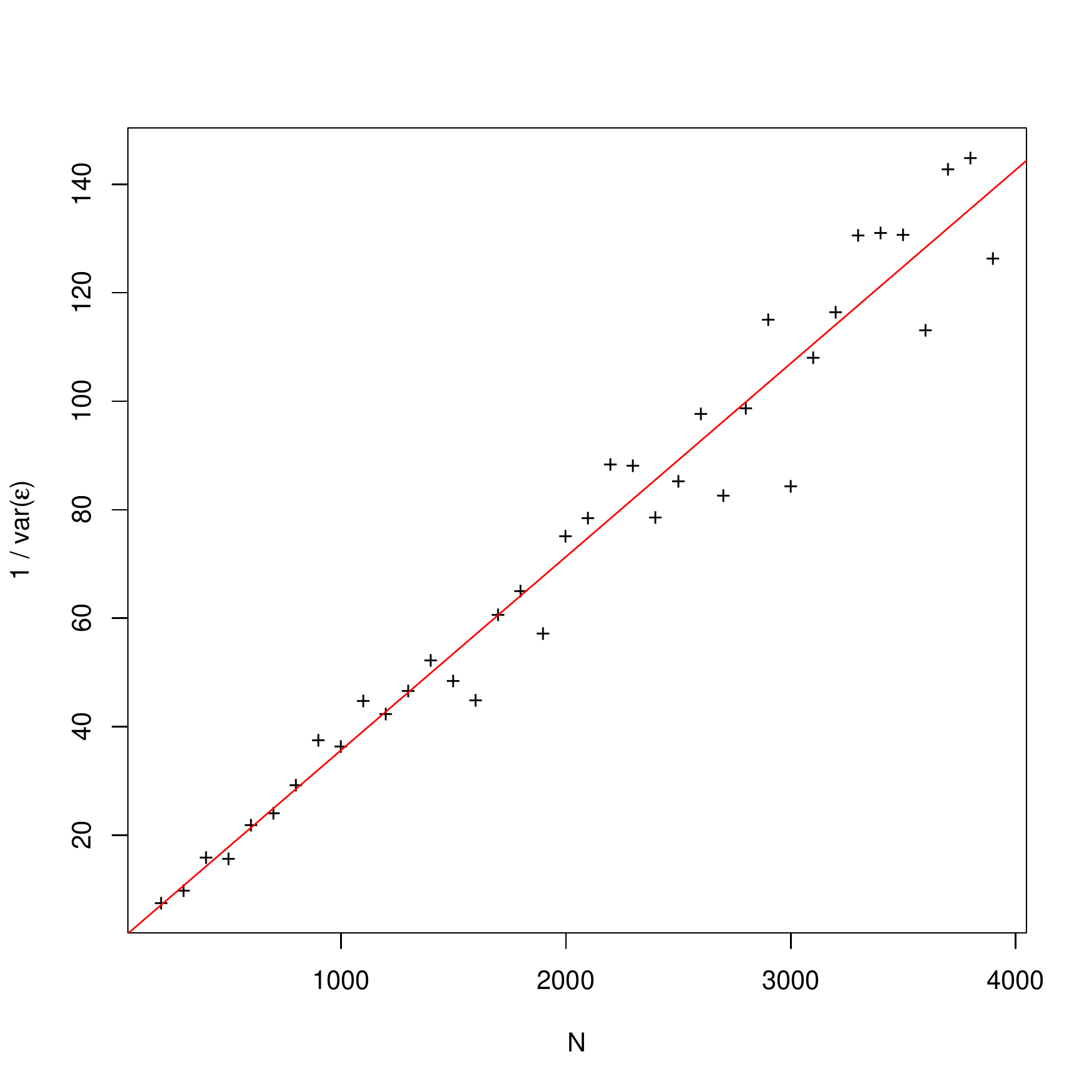}
    \caption{\label{fig.1} \textbf{Numerical stability of the
     calibration obtained after the $1$st iteration.} \emph{Top:} $0.95\%$ intervals for the calibrated
     tolerance level $\eps_2$. \emph{Bottom:} inverse of $\eps_2$'s variance.
     Variances and confidence intervals were computed on 250
     independent replicates. }
\end{figure}
\begin{figure}[n]
    \includegraphics[width=0.45\linewidth]{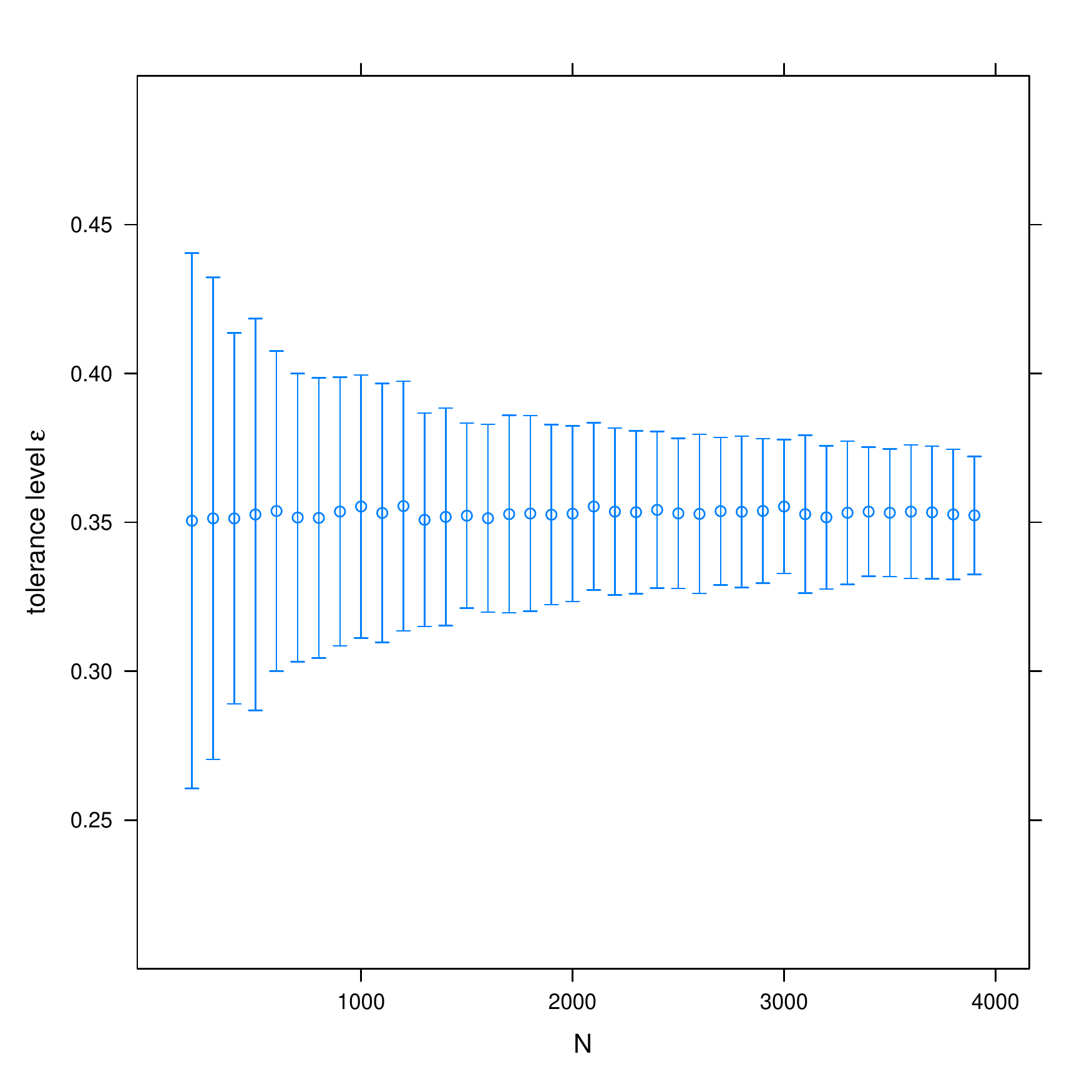}
    \includegraphics[width=0.45\linewidth]{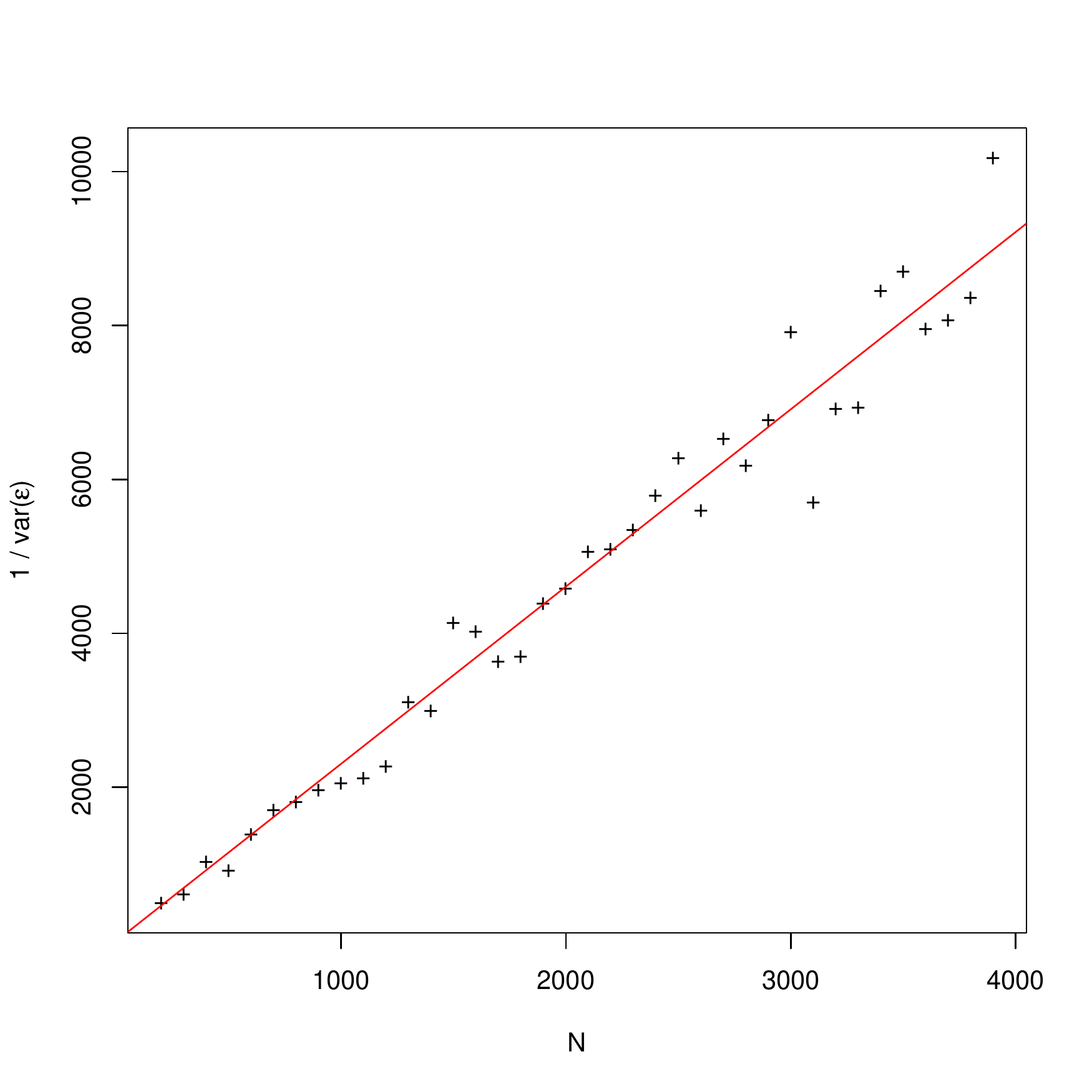}
    \caption{\label{fig.2} \textbf{Numerical stability of the
        calibration obtained after the $8$th iteration.} {\em Top}: $0.95\%$ intervals for the calibrated tolerance level
      $\eps_9$. {\em Bottom}: inverse of $\eps_9$'s variance. Variances
      and confidence intervals were computed on 250 independent
      replicates. }
\end{figure}

\noindent The following toy example, due to \citet{Sisson:Fan:Tanaka:2007},
consists of a simple mixture of two univariate Gaussian distributions
with unknown common mean $\bt$ and variances equal to $1$ and $0.01$
respectively:
\begin{equation}
  \label{eq:SISSON_MODEL}
  f\big(\bx|\btheta\big)= \varphi(\bx-\btheta)/2 
  + 10\varphi\big(10(\bx-\btheta)\big)/2
\end{equation}
The prior distribution on $\bt$ is uniformly distributed over the
interval $[-10,10]$. For the observation $\bxobs=0$, the posterior is 
\begin{equation}
  \label{eq:SISSON_POSTERIOR}
  \pi(\btheta | \bxobs=0) \propto \big[\varphi(\btheta) 
+ 10\varphi\big(10\,\btheta)\big]
\mathbf 1\big\{ -10 \le \bt \le 10 \big\}.
\end{equation}
This example is simple enough to compare sampling schemes among them
and with the true posterior.
\\

\noindent{\bf $\triangleright{}$ Is the ESS a good assessement of
quality?} We now have to test whether or not the accuracy of the output of
an algorithm with an ESS of $n_\text{final}$ is comparable to a
perfect sample of size $n_\text{final}$ drawn from the target. For
this purpose, on $200$ independent runs of the efficient scheme, we
evaluated the errors resulting from using the output of our algorithm
to evaluate functionals of the target. We compare those with
the errors associated with perfect samples.  Fig.~\ref{fig:good.ESS} studies four
functionals of the target: the mean, the median, the $1$st and the
$3$rd quartiles. In all cases, the differences on the error of both
schemes is much smaller than the differences on the error of
independent replicates of both schemes.  There is almost no
dissimilarity for estimates of the median and of $1$st and $3$rd
quartiles. We note that the interval bounded by these last
two statistics provides an ABC approximation of the credible interval
with probability $1/2$, relevance of estimating them
correctly.

\begin{table}
  \caption{\bf Comparison of ABC samplers}
  {
    \centering
    \begin{tabular}{lcc}
      \hline\noalign{\smallskip}
      & \textbf{Cost} & \textbf{ESS} 
      \\
      \noalign{\smallskip}\hline\noalign{\smallskip}
      % Accept-reject                    & $37 \times 10^5$  & $33285$    \\
      \citet{Drovandi:Pettitt:2011}      & $109\times 10^5$  & $32000$    \\
      \citet{DelMoral:Doucet:Jasra:2012} & $46 \times 10^5$  & $29250$   \\
      ABC rejection                      & $37 \times 10^5$  & $33285$    \\
      Our self-calibrated sampler        & $23 \times 10^5$  & $33285$  \\
      \noalign{\smallskip}\hline
    \end{tabular}\\[.2em] 
  }
  \footnotesize The target of the above four schemes is fixed at tolerance level
  $\epsilon=0.09$. Cost is evaluated in term of numbers of simulations from
  the model \eqref{eq:SISSON_MODEL}. As shown in the third column,
  algorithms were tuned to produce outputs with about the same quality
  in term of effective sample size (ESS). 
  \label{tab:toy}
\end{table}

\noindent{\bf $\triangleright{}$ Comparison with other samplers.}
In Tab.~\ref{tab:toy}, we compare four algorithms, namely those of
\citet{Drovandi:Pettitt:2011}, \citet{DelMoral:Doucet:Jasra:2012} when
using $M=1$ data set per particles, the rejection sampler
(Alg.~\ref{algo:basic}) and our proposal.  In all cases, the target
has tolerance level $\epsilon=0.09$. Our proposal was tuned with $N=
10^5$ particles at the sequential stage. The other samplers were
tuned to produce a sample with the same accuracy in term of ESS and
tolerance level. The scheme of
\citet{DelMoral:Doucet:Jasra:2012} was disavantaged by attaching only
$M=1$ simulated data set per particles, since their calibration was
designed for larger values of $M$. As seen from Tab. \label{tab:toy},
our sampler needs to reach the required accuracy less simulations
from the model than the other three ones.

\noindent{\bf $\triangleright{}$ Stability of the calibration.}
We also study the numerical stability of the calibration scheme based on
250 independent replications of the whole algorithm, when the tuning
parameter $N$ (ie the size of the array during the sequential stage) varies
from $200$ to $4000$. Fig.~\ref{fig.1} and \ref{fig.2} detail the
stability of the $1$st and $8$th iteration. They indicate that
the variance of the calibrated tolerance levels is of order $1/N$. Here,
values of $N$ are artificially small to allow us to perform 250
replicates of the whole scheme. The typical values of $N$ we had in
mind when designing the algorithm are around $10\,000$ or
$100\,000$. Results of Fig.~\ref{fig.1} and \ref{fig.2} indicate that
the calibrated $\epsilon_t$'s have a tendency to converge to a fixed
value with $N$, with a mean square error of order $1/N$.

\begin{figure}[n]
  \begin{tabular}{cc}
    \includegraphics[width=.45\linewidth]{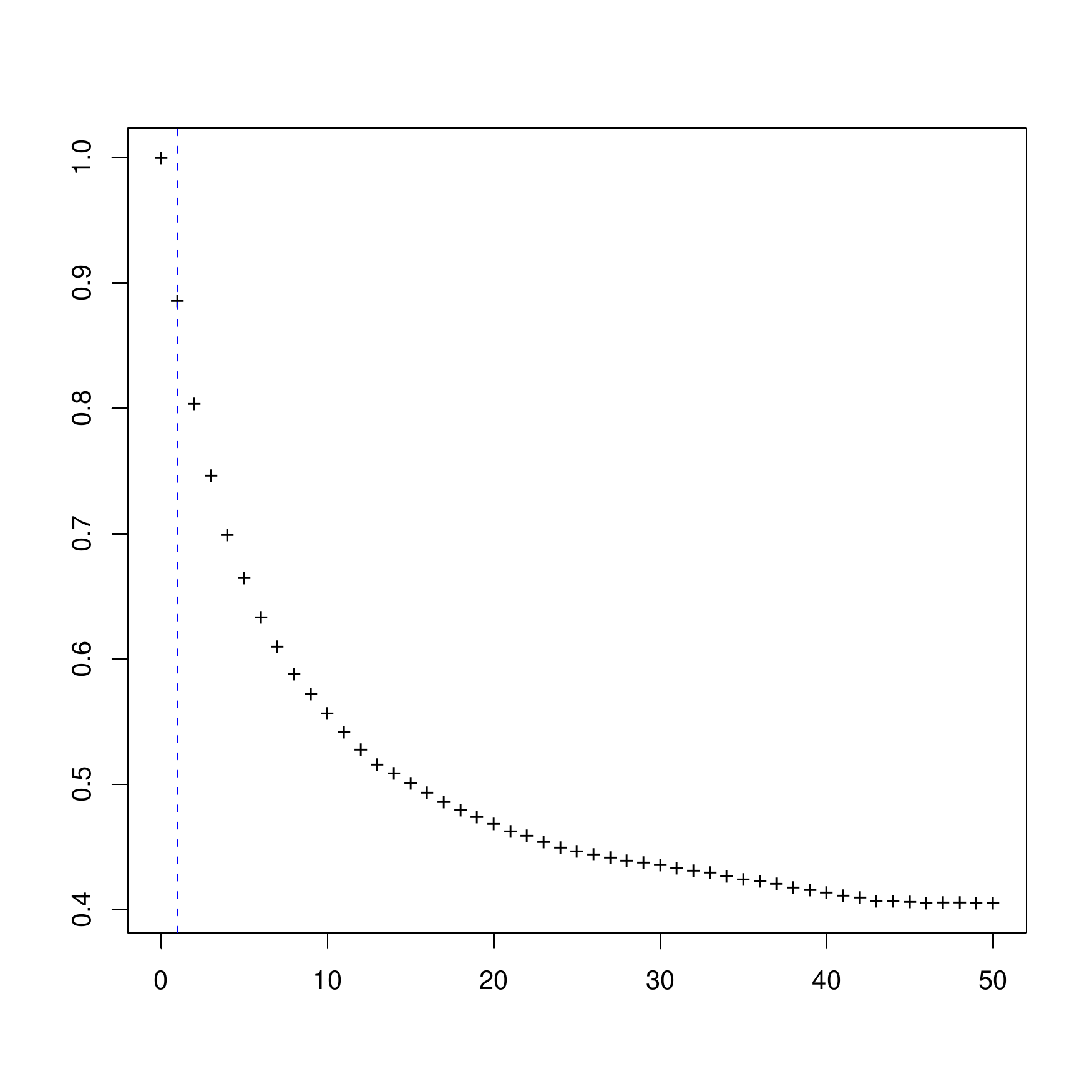}&
    \includegraphics[width=.45\linewidth]{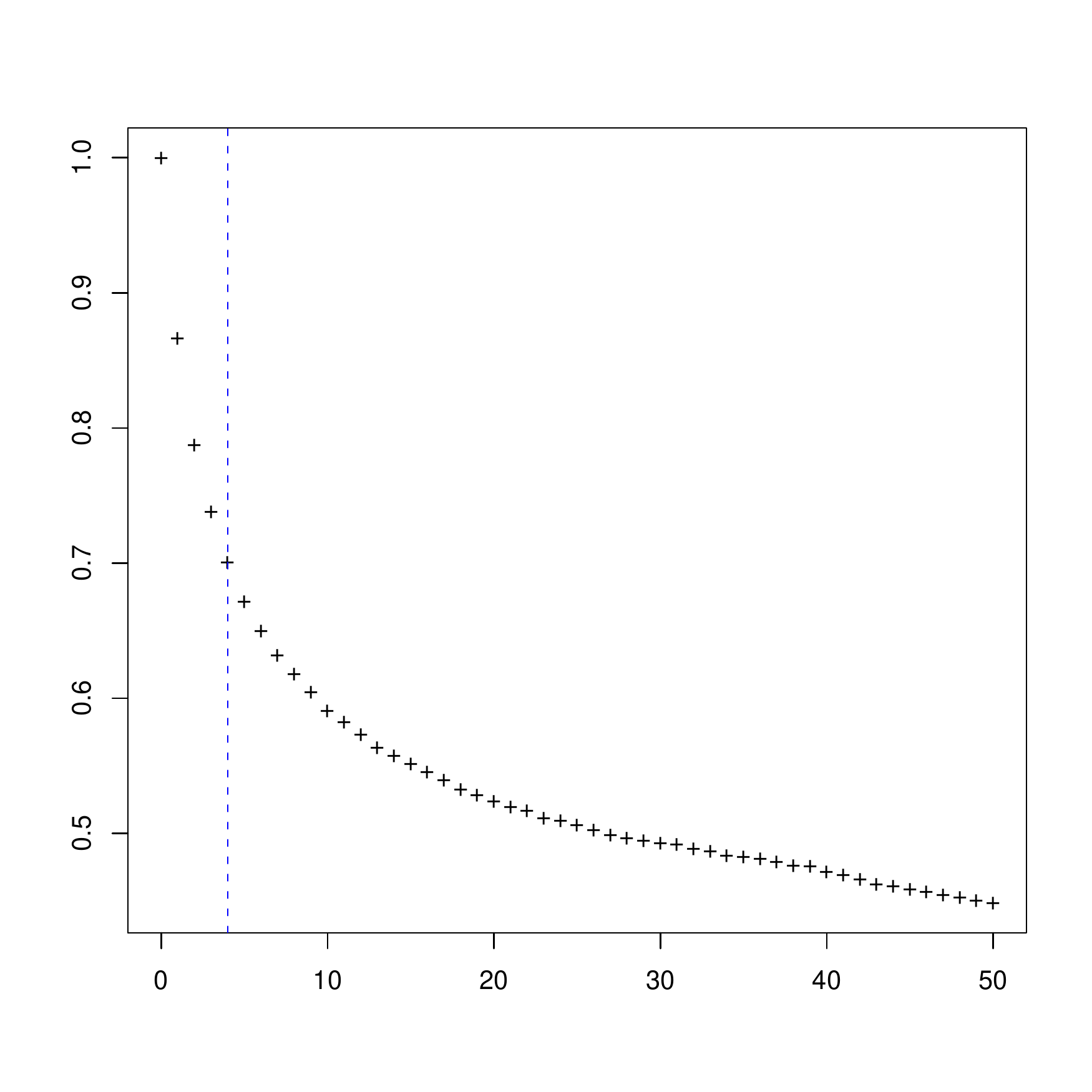}\\
     \includegraphics[width=.45\linewidth]{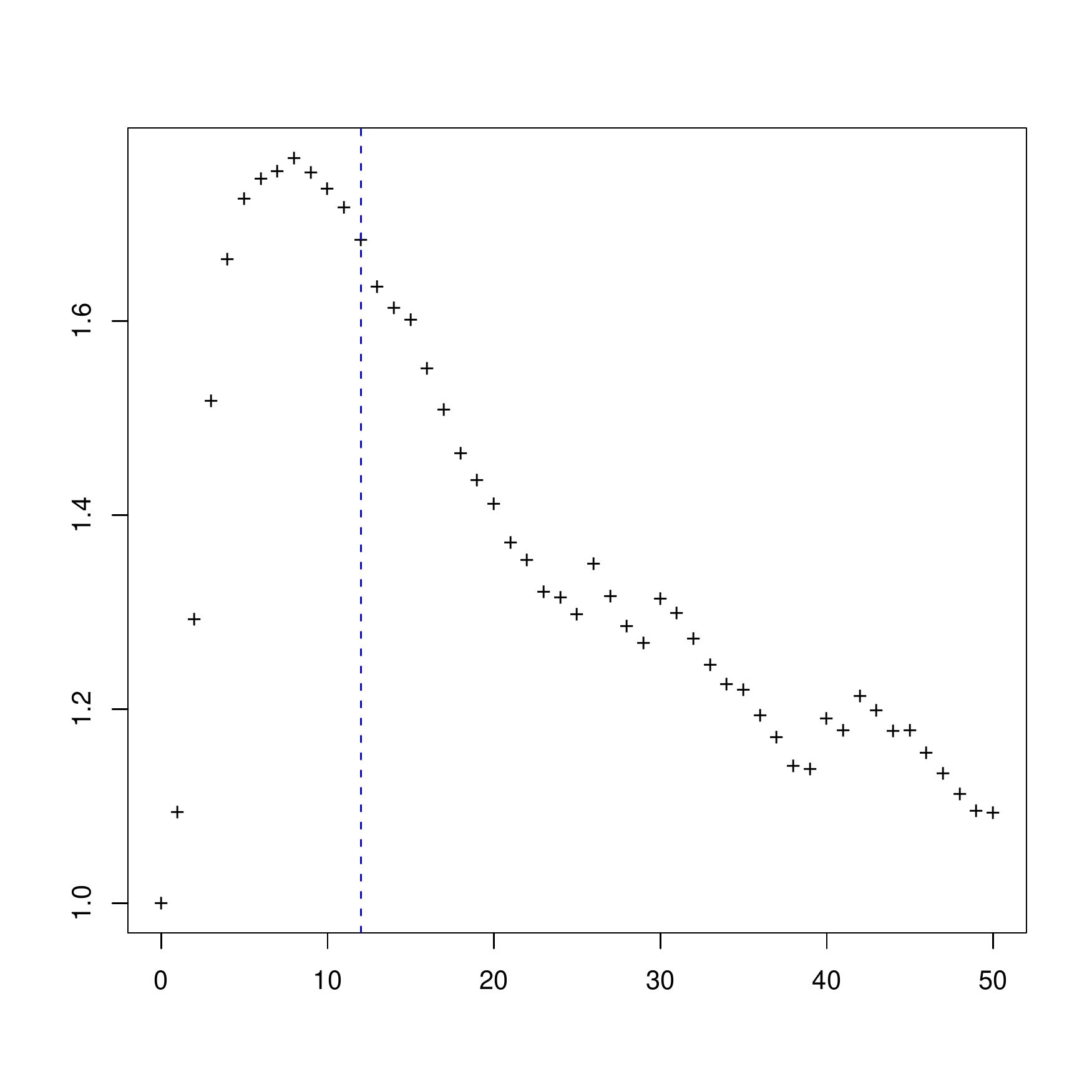}&
    \includegraphics[width=.45\linewidth]{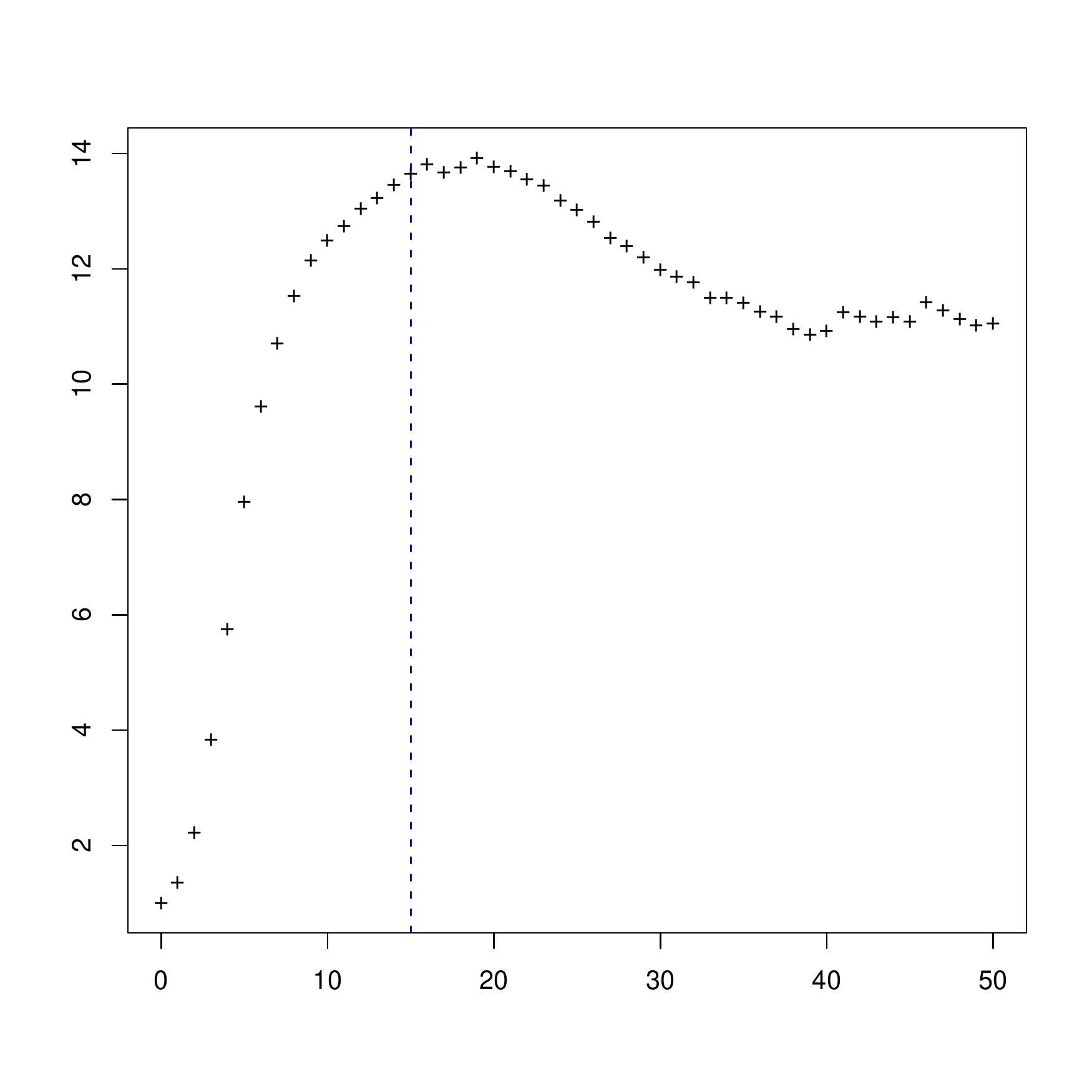}
    \end{tabular}
    \caption{\label{fig:ABC:3} {\bf Efficiency of our self calibrated scheme.}  In all
    plots, $x$-axis: number of iteration and $y$-axis: the gain factor defined in
    Paragraph~\ref{sec:measurements}. The stopping times given by the criterion described in
    Paragraph~\ref{sub:stop} are represented by the dotted vertical lines.
    \emph{Top, left:} uniform prior on $[-0.1,0.1]$ and
    $\alpha_0 = 1/10$.  \emph{Top, right:} uniform prior on $[-1; 1]$ and $\alpha_0 = 1/10$.
    \emph{Bottom, left:} uniform prior on $[-10,10]$ and $\alpha_0 = 1/2$.
    \emph{Bottom, right:} uniform prior on $[-100,100]$ and $\alpha_0 = 1/4$.}
\end{figure}

\noindent{\bf $\triangleright{}$ Efficiency with respect to the
  gap between the prior and the posterior.}
We mentioned earlier (in Remark~\ref{rem:learn.SMC} and at the end of
Paragraph~\ref{sub:init}) that sequential schemes are not recommended
when the prior provides sufficient information about the parameter.
The four curves in Fig.~\ref{fig:ABC:3} correspond to various priors
on the parameter $\btheta$ in Eq.~\eqref{eq:SISSON_MODEL}. They show the
evolution of the gain factor introduced in \eqref{eq:gain_factor} over
the iterations. The vertical dotted line (blue) indicates the stopping
time given by the criterion of Paragraph~\ref{sub:stop}.  The top two
curves of Fig.~\ref{fig:ABC:3} correspond to the two most informative
priors. In such cases, the stopping criterion is met in the early
iterations and the efficiency is not in favor of sequential schemes.
In each of the four studies, we tuned the initialisation stages to end
with values of $\alpha_0$ equal to $1/10$, $1/10$, $1/2$, $1/4$
respectively. The two bottom curves in Fig.~\ref{fig:ABC:3} indicate
that, at least on this toy example, the efficiency of our proposal
increases with the difference between prior and posterior. Thus, we
conclude that our scheme can significantly save time (over the
rejection algorithm) when the prior provides little information
compared to the information carried by the data.

\subsection{A population genetics example}\label{sec:popgen}

\begin{figure}[n]
\centering
\includegraphics[width=0.6\linewidth]{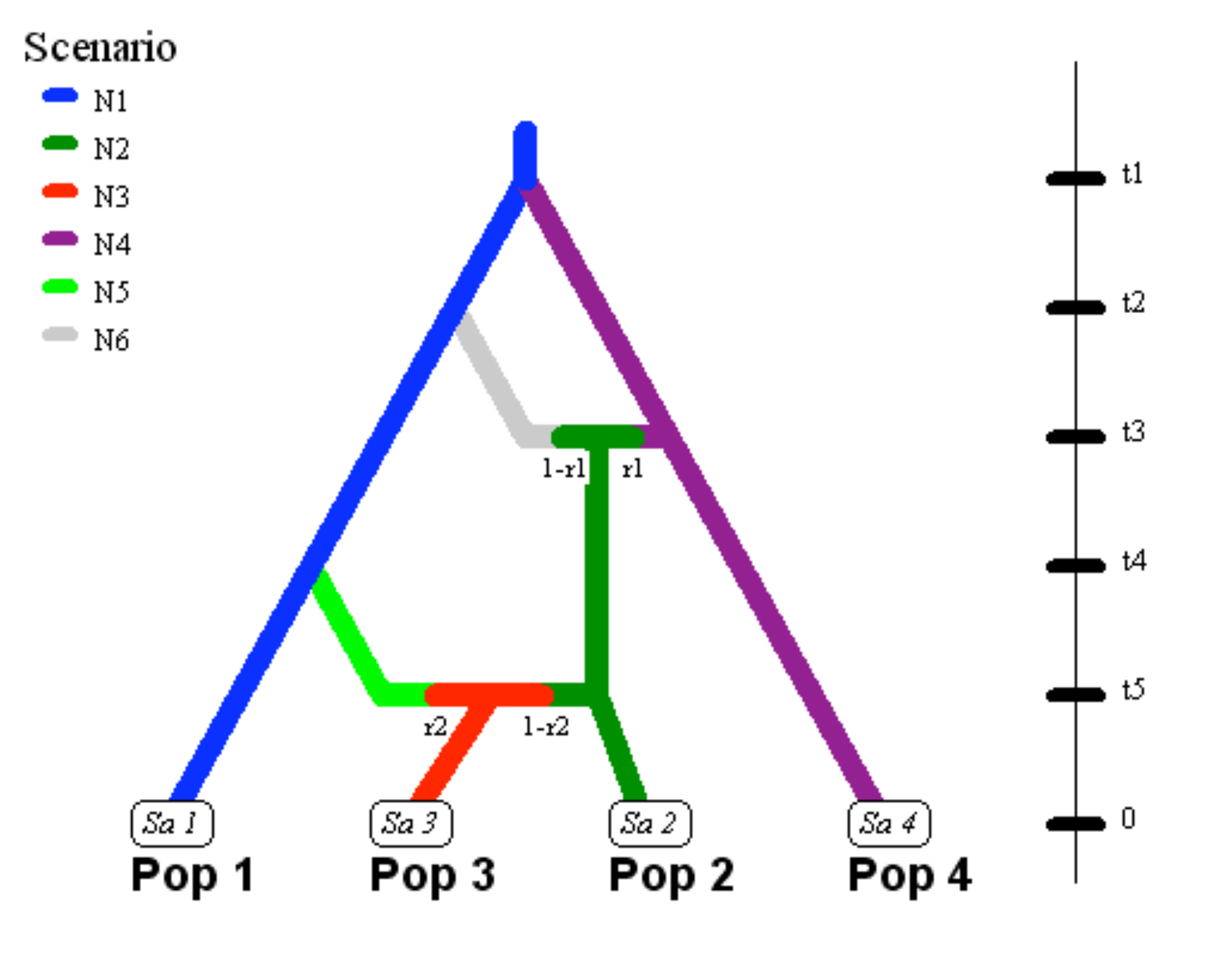}
\caption{\textbf{Evolutionary scenario} of {\it Apis
  mellifera}. Populations 1, 2, 3 and 4 are respectively in (1) the
  West coast of France, (2)  Italy, (3) Aosta Valley in the
  Alps and (4) Croatia. Populations 1, 2 and 4 represents respectively
  subspecies \textit{A. m. mellifera}, \textit{A. m. ligustica} and
  \textit{A. m. carnica}.  Meanwhile Population 3, in the area of \textit{A. m. mellifera},
  had been mixed recently with \textit{A.m. ligustica} by beekeepers.}
\label{fig:honeybees:scenario}
\end{figure}

We now consider an experiment that relates more directly to the
genesis of ABC, namely population genetics.  We perform Bayesian
inference over the parameters of the evolutionary scenario
(Fig.~\ref{fig:honeybees:scenario}) of \textit{Apis mellifera} (the common bee).
The parameter of interest includes five dates of inter-populational
events, $t_1, \ldots, t_5$ (from the oldest to the newest).  It is
composed of two types of events: three divergences parameterized by
their dates $t_1, t_2,t_4$ and two admixtures parameterized by their
dates $t_3, t_5$ and rates $r_1$ and $r_2$. The old, native area of
\textit{Apis mellifera} was Asia. Those bees invaded western Europe, bypassing
the Alps going either North or South. The split between the two
lineages is modelised by the divergence at time $t_1$.  The current
population in Italy (Population 2) came from an old mixture (at time
$t_3$) from both lineages. The admixture rate $r_1$ and $r_2$ were
respectively the frequency of \textit{A.m. carnica} in Italy and
\textit{A.m. mellifera} in Aosta Valley at the foundation of the colony.

The data set is composed of present-day genetic data on samples of
individuals collected from the four populations at the bottom of the
scenario. Each sample is composed of about forty diploid individuals,
and to obtain the data, each individual has been genotyped at
eight microsatellite loci in the autosomal part of the genome.  The
eight loci might be considered as neutral (no selection pressure) and
independent.  To perform the ABC analysis, we computed $30$ summary
statistics on the genetic data, see Appendix B.

The stochastic model is as follows: a unit time represent one
generation, which is about 24 monthes for \textit{Apis mellifera}. The
genealogy of the sample at the $\ell$-th locus is distributed
according to a coalescence process, contrained by the evolutionary
scenario of Fig.~\ref{fig:honeybees:scenario}
\citep{donnelly:tavare:1995} This random genealogy is parameterised by
the five dates $t_1,\ldots, t_5$, both admixture rates $r_1, r_2$
and the effective population sizes $N_1, \ldots, N_6$ over the
branches of the scenario as given in
Fig.~\ref{fig:honeybees:scenario}. Conditionally on that genealogy,
the microsattelite marker evolves along the branches of that tree
according to the generalised stepwise model
\citep{estoup:etal:2002}. This mutational process on the length of the
marker is the superimposition of two independent models: the first
model is composed of single nucleotide indels that occur at rate
$\mu_{SNI, \ell}$ and the second model is composed of insertion or
deletion of a random number, say $G$, of the repeated motif of that
locus, and occurs at rate $\mu_{MIC, \ell}$. The random variable $G$
follows a geometrical distribution with parameter $0.42$.

\noindent\textbf{$\triangleright$ Prior.} We have built a hierarchical
model on the parameters of the model. The parameter of interest (we
denote it by $\phi$ to avoid confusion with the fact that the letter
$\bt$ is reserved for another use in population genetics) is made
of eigth parameters: $\phi=(\widetilde N, t_1, t_2, t_3, t_4, t_5,
r_1, r_2)$.

\begin{itemize}
\item The effective population sizes
  are managed by the hyper-parameter $\widetilde{N}$ with a uniform
  prior over $[5,100\,000]$. The six population sizes $N_k$,
  $k=1,\ldots,6$, were drawn independently from a Gamma distribution with
  position and shape respectively equal to $\widetilde{N}/2$ and $2$, and
  truncated to be in $[1,500\,000]$.
\item The admixture rates $r_1$ and $r_2$ have independent uniform
  priors on $[0.01,0.999]$.
\item The prior on the six event dates ensures that $t_1\ge
  t_2\ge t_3,\ t_4\ge t_5$ and is given by:
  \begin{align*}
  t_5 &\sim  \mathcal U_{[0.01,800]}\,, 
  \\
  (t_4 - t_5) &\sim \mathcal U_{[0,50\,000]}\,,
  \\
  (t_3 - t_5) &\sim \mathcal U_{[0,50\,000]}\,,
  \\
  t_2 - \max\big(t_3, t_4\big) &\sim \mathcal U_{[100,500\,000]}\,, 
  \\
  (t_1 - t_2) &\sim \mathcal U_{[100,2\, 000\, 000]}\,.
  \end{align*}
\item At last, the mutation model is managed by fixed
  hyper-parameters: $\widetilde{\mu}_{SNI}=10^{-6}$ and
  $\widetilde{\mu}_{MIC}=10^{-4}$. The rates at the $\ell$-th locus are
  then drawn as
  \begin{align*}
    \mu_{SNI, \ell}& \sim {\Gamma}_\text{trunc}(
    \widetilde{\mu}_{SNI}\big/ 2,\  2,\ 10^{-9},\ 10^{-3}),
    \\
     \mu_{MIC, \ell}& \sim {\Gamma}_\text{trunc}(
    \widetilde{\mu}_{MIC}\big/ 2,\  2,\ 5\times 10^{-7},\ 5\times 10^{-2}),
  \end{align*}
  where $\Gamma_\text{trunc}(p,\ s,\ a,\ b)$ is the Gamma distribution
  with position $p$, shape $s$ truncated on the interval
  $[a;b)$.
\end{itemize}

\begin{figure}[n]
\centering
\includegraphics[width=0.45\linewidth]{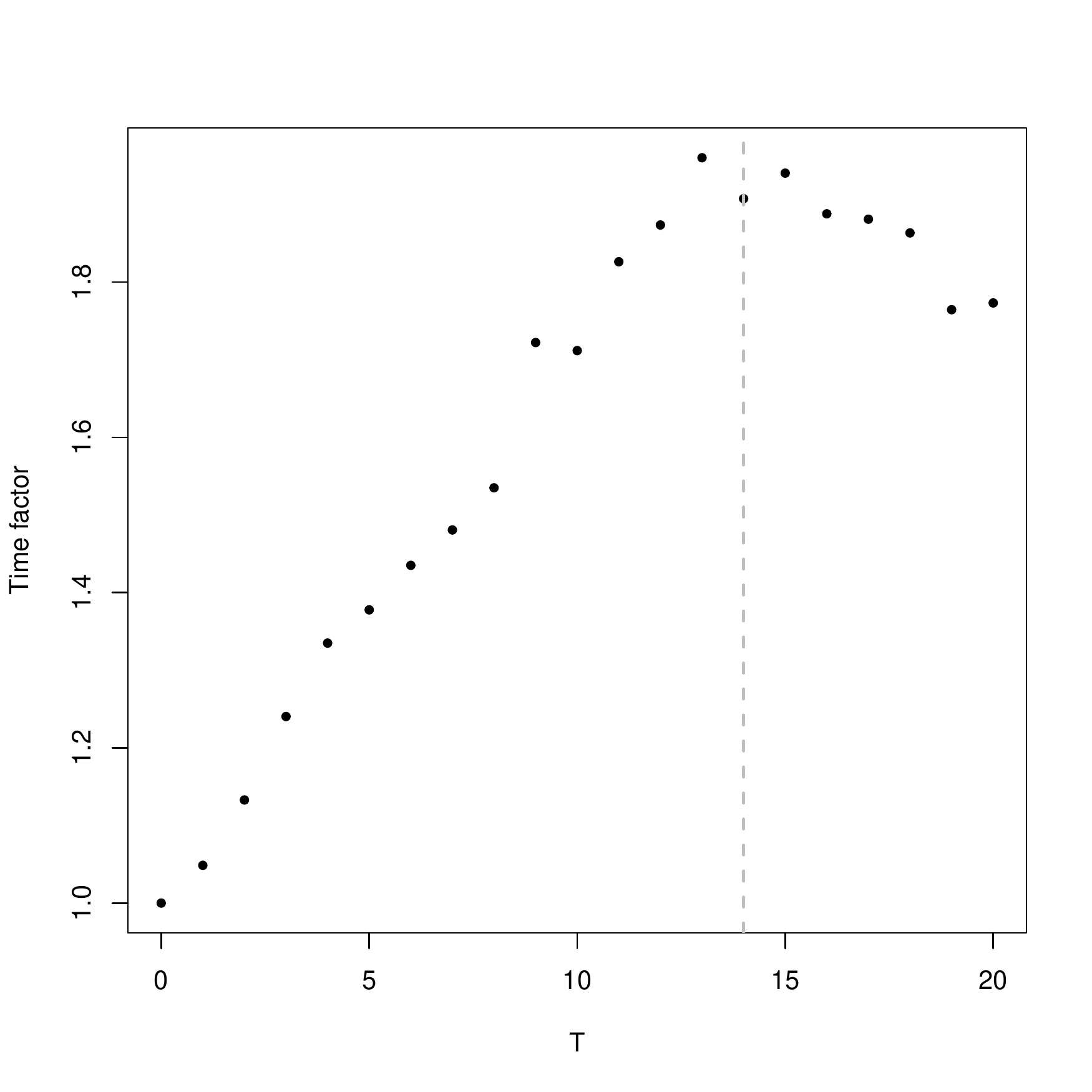}
\caption{{\bf Efficiency of our self-calmibrated scheme} estimated
the gain factor in the population genetics example.}
\label{fig:honeybees:timefactor} 
\end{figure}

% Fig.\ref{fig:honeybees:thetas} presents the results for the estimates
% of the date parameters.  One can see the stability of the
% posterior density estimate at the end of the proposed algorithm on
% five independent replicates of the whole scheme. 

\begin{figure}[n]
  \begin{center}
    \includegraphics[width=0.8\linewidth]{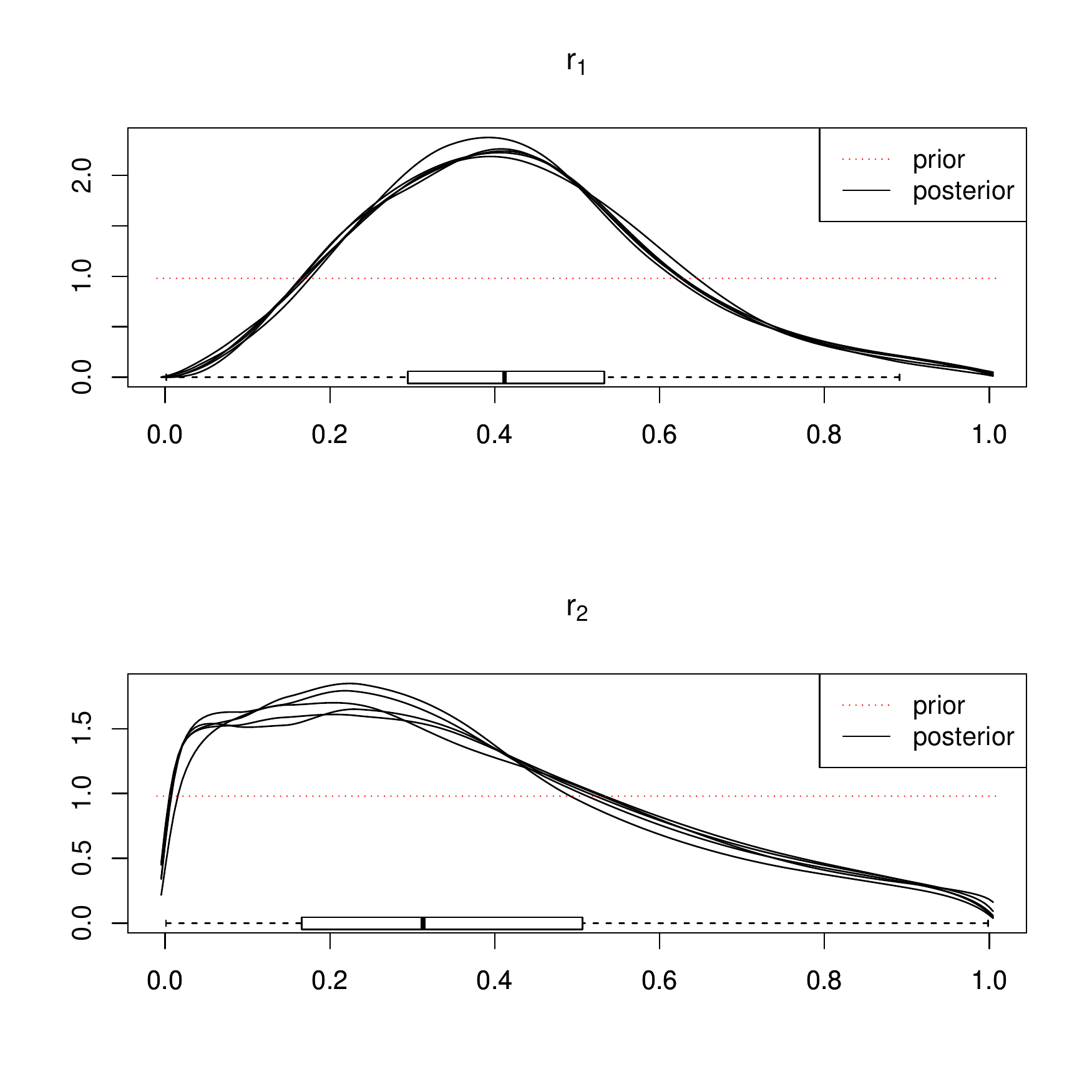}
    \caption{%
      {\bf Stability of the estimated posterior densities} for the two admixture
      rates on five independent replicates of our self-calibrated scheme.
      \label{fig:honeybees:admixtures}
}
\end{center}
\end{figure}

Fig.~\ref{fig:honeybees:admixtures} shows the ABC posterior of both
admixture rates. Finally, the gain factor of our proposal compared to
the standard ABC rejection agorithm is given in Figure
\ref{fig:honeybees:timefactor}. The dashed vertical line indicates the
stopping time of the iterative algorithm (\textit{i.e.}, the first $T$
for which $\rho_T\le 0.1$) that is equal to $14$ iterations in the
five independent replicates. We only need half the number of
simulations from the model to obtain results similar to those of
the standard ABC accept-reject algorithm.

% \begin{figure*}[n]
%   \centering
%   \includegraphics[width=\textwidth]{theta-abeilles.pdf}
%   \caption{%
%     {\bf Stability of density estimates of the posterior
%       distribution} of the date parameters on five independent
%     replicates of the proposed scheme. XXX THE ABOVE FIGURE IS NOT THE
%     CORRECT ONE!!! XXX
%     \label{fig:honeybees:thetas} 
%   }
% \end{figure*}

\section{Discussion}
\label{sec:discussion}
The paper proposes a likelihood-free, sequential algorithm calibrated
in such a way as to use a relatively small number of simulations from the
model while maintaining a suitable level of accuracy. In the complex
example from population genetics, the SMC algorithm is about twice
faster than the simple rejection algorithm. Present advances in
microbiology increase the number of loci in the genetic data
sets, and then the amount of information in the data, but slow
down generating simulated data sets with the same number of loci in
ABC algorithms.  For instance, with the next generation sequencing (NGS),
geneticians should be able to produce thousands of SNPs' loci. An
efficient self-calibrated sampler is a promising way to treat such
large data sets if the number of summary statistics does not increase.

Throughout the paper we have left the size $N$ of the array in
the sequential stage as a free parameter but it could be helpful
for the user to provide guidelines on how to tune that parameter.
Actually, the size should depend on the value of the final level
$\epsilon_T$ and its distance from the tolerance level $\epsilon$ of
the final target. If $\epsilon_T$ is not small enough, the rejection
in the post-processing stage fixes the size of the very final output,
which can be drastically lower than $N$. Because neither $T$ nor
$\epsilon_T$ are known in advance, tuning $N$ is a difficult question.
Our recommendation is to perform a preliminary but complete run of the
whole algorithm with a value of $N$ intentionally too small (less than
a thousand) to get rough estimates of $T$ and $\epsilon_T$ and restart
the algorithm with a suitable value of $N$.

In addition, we did not provide a formal proof that our calibrating
scheme do not bias the na\"{i}ve sequential Monte Carlo scheme, whose
correctness is already given in the literature. But we have
tried to provide informal arguments in favor of our algorithm, and
tested them in numerical experiments.  A formal theoretical
study of the proposal seems like a major challenge to us, because of
the resampling steps and estimation of additional quantities for
calibrating the tolerance levels.

Other important issues were set aside in this paper, like constructing or
choosing the summary statistics, tuning the final tolerance
level. For instance, one could adapt the
efficient sequential algorithm to include an automatic construction of
the summary statistics as in \citet{fearnhead:prangle:2012}.
At last, the efficient scheme deals with parameter estimation, and
not with model choice. Whether or not we can provide an efficient sampler to perform a
Bayesian model choice, inspired from the proposal of the paper, is
still on open question.

\section*{Acknowledgements} 

We would like to warmly thank the Associate Editor and both reviewers for their comments and
suggestions. They have helped us to improve the paper both in the presentation and in the
scientific content.  All four authors have been partly supported by the \textit{Agence Nationale de la
Recherche} (ANR) project EMILE (ANR-09-BLAN-0145-04).  The $2$nd and $4$rd authors have also been
supported by the \textit{Labex} NUMEV (\textit{Solutions Num\'eriques, Mat\'erielles et Mod\'elisation pour
l'Environnement et le Vivant}).

\bibliographystyle{apalike}
\bibliography{SCMPR13} 

\appendix
\section*{Appendix A -- Proof of the approximation in
  Eq.~\eqref{eq:n_t}}

On average, the number of different particles after the resampling is
$\alpha_t\esp(n_t)$. The Markov evolution of the copies might add
new particles, and remove old ones. The Markov move is applied
to each particle of the array of size $N$, and displace them with
probability $\rho_t$. Thus, on average, the Markov evolution add
$\rho_tN$ particles. 

Moreover, old particles might disappear if all
their copies are displaced by the Markov kernel. The probability that
this occur depends on the number of copies. This number is always
larger than $\ell_t:=N/(\alpha_t N)$ because each kept particle is
copied $\ell_t$ times during the last residual resampling.  Hence an
old (but kept) particle is removed with probability smaller than
$\rho_t^{\ell_t}$, and the expected number of old particles that
disappear because of the Markov moves might then be bounded from
below by $\rho_t^{\ell_t}\alpha_t\esp(n_t)=o(\alpha_t)\esp(n_t)$. Morover,
$\rho_t^{\ell_t}$ is negligible in front of the other terms because
$\alpha_t \ll 1$, \textit{i.e.}, $\ell_t\gg 1$.  Hence,
\begin{equation*}
\esp(n_{t+1}) 
 \ge (\alpha_t + \rho_t + o(\alpha_t))\esp(n_t).
\end{equation*}

\section*{Appendix B -- Summary statistics of the %
  population genetic experiment}
Sixteen statistics are computed within
each sample: the average (over each locus) of the numbers of alleles,
the average of the variances of the allele lengths, the average of the
heterozygosity, the average of the M index of Garza and Williamson. Twelve
statistics are computed on each pair of samples: the average of
the Weir and Cockerham's $F_{st}$ and the average of the
$(\delta\mu)^2$ of Goldstein. The last two statistics are
a rough estimate of the frequency of individuals in Population 3 that are
most likely to come from Population 1 than Population 2, and a rough
estimate of the frequency of individuals in Population 2 that are
most likely to come from Population 4 than Population 1. Thoses
statistics were computed as in the DIYABC software of
\cite{Cornuet:etal:2008}.

\end{document}